\documentclass[pra,twocolumn,superscriptaddress,notitlepage,showpacs,showkeys]{revtex4-1}
\usepackage{graphicx,amsmath,amsfonts,amssymb,upgreek,txfonts,color,mathtools,braket,soul}
\usepackage[colorlinks,linkcolor=blue,citecolor=blue,urlcolor=blue,breaklinks=true]{hyperref}
\usepackage[utf8x]{inputenc}

\begin{document}
	
\title{Homodyne coherent quantum noise cancellation in a hybrid optomechanical force sensor}

\author{H. Allahverdi}
\address{Laser and Plasma Research Institute, Shahid Beheshti University, Tehran, Tehran 19839-69411, Iran}

\author{Ali Motazedifard}
\email{motazedifard.ali@gmail.com}
\address{Department of Physics, University of Isfahan, Hezar-Jerib, Isfahan 81746-73441, Iran}
\address{Quantum Optics Group, Department of Physics, University of Isfahan, Hezar-Jerib, Isfahan 81746-73441, Iran}
\address{Quantum Sensing Lab, Quantum Metrology Group, Iranian Center for Quantum Technologies (ICQT), Tehran, Tehran 15998-14713, Iran}

\author{A. Dalafi}
\address{Laser and Plasma Research Institute, Shahid Beheshti University, Tehran, Tehran 19839-69411, Iran}

\author{D. Vitali}
\address{Physics Division, School of Science and Technology, University of Camerino, 62032 Camerino, Macerata, Italy}
\address{INFN, Sezione di Perugia, Via A. Pascoli, 06123 Perugia, Perugia, Italy}
\address{CNR-INO, Largo Enrico Fermi 6, 50125 Firenze, Italy}

\author{M. H. Naderi}
\address{Department of Physics, University of Isfahan, Hezar-Jerib, Isfahan 81746-73441, Iran}
\address{Quantum Optics Group, Department of Physics, University of Isfahan, Hezar-Jerib, Isfahan 81746-73441, Iran}

\date{August 11, 2022}

\begin{abstract}
In this paper we propose an experimentally viable scheme to enhance the sensitivity of force detection in a hybrid optomechanical setup assisted by squeezed vacuum injection, beyond the standard quantum limit (SQL).
The scheme is based on a combination of the coherent quantum noise cancellation (CQNC) strategy with a variational homodyne detection of the cavity output spectrum in which the phase of the local oscillator is optimized. In CQNC, realizing a negative-mass oscillator in the system leads to exact cancellation of the backaction noise from the mechanics due to destructive quantum interference. Squeezed vacuum injection enhances this cancellation and allows sub-SQL sensitivity to be reached in a wide frequency band and at much lower input laser powers. We show here that the adoption of variational homodyne readout enables us to enhance this noise cancellation up to $40~\mathrm{dB}$ compared to the standard case of detection of the optical output phase quadrature, leading to a remarkable force sensitivity of the order of $10^{−19}~\mathrm{N}/\sqrt{\mathrm{Hz}}$, about $70\%$ enhancement compared to the standard case. Moreover, we show that at nonzero cavity detuning, the signal response can be amplified at a level three to five times larger than that in the standard case without variational homodyne readout, improving the signal-to-noise ratio. Finally, the variational readout CQNC developed in this paper may be applied to other optomechanical-like platforms such as levitated systems and multimode optomechanical arrays or crystals as well as Josephson-based optomechanical systems.
\end{abstract}
	
	
\maketitle

\section{Introduction}

Optomechanical systems (OMSs) have been exploited in different areas of quantum technologies, such as quantum information processing and communication
\cite{stannigel2012optomechanical,stannigel2011optomechanical,stannigel2010optomechanical,rogers2014hybrid}, quantum memories \cite{sete2015high,wallucks2020quantum},
reversible microwave-to-optics converters \cite{andrews2014bidirectional,forsch2020microwave,jiang2020efficient,lauk2020perspectives,arnold2020converting}, microwave
circulators \cite{barzanjeh2017mechanical,shen2018reconfigurable}, quantum correlations
\cite{foroudcrystalentanglement,polzikDistantEntanglementOMS2020,foroudsynch,barzanjehentanglement,dalafiQOC}, and quantum squeezing \cite{giovanniCQNC2020,aliDCEsqueezing}, as
well as in fundamental physics \cite{teleportation,optomechanicalBelltest1,Pikovski2012, PhysRevLett.116.070405, PhysRevLett.107.020405,aliDCE2, aliDCE3,NoriDCE1,aliGreen2021,Heforce2022}.
Also, optomechanical-based sensors have been recognized as an optimal candidate for the detection of minuscule forces at the quantum limit \cite{braginsky1995quantum} such as
observation of gravitational waves \cite{virgoBAexperiment2020}.
In OMSs, cavity field shot noise and radiation pressure backaction (BA) noise restrict the force measurement sensitivity, leading to the standard quantum limit (SQL) in force
measurements \cite{aspelmeyer2014cavity,meystre2013short}.
These two noise sources show opposite responses to the input field power: As the cavity driving power grows, the shot noise decreases, while the radiation pressure BA noise
increases \cite{aspelmeyer2014cavity}. Thus, lowering one results in strengthening the other one. Consequently, at high input powers where quantum effects are enhanced
and the shot noise is negligible, the mechanical BA noise is dominant. Therefore, any effort to improve the measurement sensitivity requires suppression of the BA noise.

Various strategies have been proposed for suppressing the BA noise effects to achieve an ultrasensitive measurement using optomechanical systems, such as utilizing a shot-noise-limited microwave interferometer
\cite{teufel2009nanomechanical}, degenerate parametric amplification \cite{huang2017robust}, back-action evasion techniques \cite{chao2022backaction,fani2020back, clerk2008back, sillanpaa2020ForceFree}, variational readout of the cavity output field \cite{Kampel2017improve}, a measurement-based feedback technique \cite{bemani2021force}, coherent quantum noise cancellation (CQNC) \cite{tsang2010coherent}, and a noise reduction scheme in systems with a single-mode mechanical resonator in the hybrid system \cite{gong2021weak,lee2020squeezed,aliDCEforcesenning,mehryMagneticsensing2020,Jong2022cooperativity} as well as in systems with multi-mode mechanical oscillators \cite{Yanay2016, Burgwal2019, Dumont2022, Paraiso2015}.
Among them, CQNC is one of the most successful approaches of noise suppression
which utilizes quantum interference. The idea of CQNC, which was first introduced by Tsang and Caves~\cite{tsang2010coherent}, is based on the exploitation of an ancillary
oscillator with an effective negative mass to create an antinoise path to the system dynamics to cancel out the BA noise of the mechanics. A scheme based on the
CQNC method in a hybrid OMS consisting of an atomic ensemble and equipped with squeezed vacuum injection has been proposed in Refs.~\cite{bariani2015atom,motazedifard2016force,motazedifard2021ultraprecision}, and it has been recently realized experimentally~\cite{cQNCNatureexp}. By calculation of the output cavity phase
spectrum, it has been shown that under the so-called perfect CQNC conditions, the BA noises due to the coupling of the intracavity radiation pressure with the mechanical
oscillator (MO) and with the atomic ensemble completely cancel each other. Furthermore, it has been demonstrated that the cavity shot noise can be suppressed by injecting a
squeezed vacuum light into the cavity. These proposals rely on measuring the phase quadrature spectrum of the output cavity field.


Motivated by the above-mentioned investigations, we have been encouraged to improve the output detection method
by exploiting a variational homodyne readout \cite{Kimble2001} (see also Refs.~\cite{Kampel2017improve,Mason2019}) in which the local oscillator phase is optimized for each parameter set rather than
considering the standard quadrature phase measurement common in optomechanical force sensing.
Under perfect CQNC conditions, which refers to zero cavity detuning (resonant case) and perfect matching between positive- and negative-mass MO parameters,
optimizing the phase of the homodyne detected quadrature has no advantage. Nevertheless, for nonzero detuning and/or in the presence of a mismatch between
mechanical and atomic parameters, properly adapting the phase of the detected quadrature enables us to achieve noise reduction compared to the
standard case of Ref.~\cite{motazedifard2016force}, up to $40~\mathrm{dB}$ when adjusting the optomechanical parameters.
Moreover, one can also amplify the cavity output signal compared to the standard CQNC scheme and, as a result, one can achieve simultaneous noise reduction and output signal amplification in some cases.

The paper is organized as follows. Section~\ref{sec2} is devoted to the description of the physical model of the system under consideration. In Sec.~\ref{sec3} the linearized
quantum Langevin equations (QLEs) describing the system dynamics are obtained. The enhancement of CQNC in the output spectrum of a generic optical quadrature is shown and discussed
under different conditions in Sec.~\ref{sec4}. The advantage of the variational homodyne CQNC setup in force sensitivity, the signal-to-noise ratio (SNR), and output signal
amplification are discussed in Sec.~\ref{sec5}. Conclusions are summarized in Sec.~\ref{sec6}.

\section{\label{sec2}The system }
\begin{figure}
	\begin{center}
		\includegraphics[width=8cm]{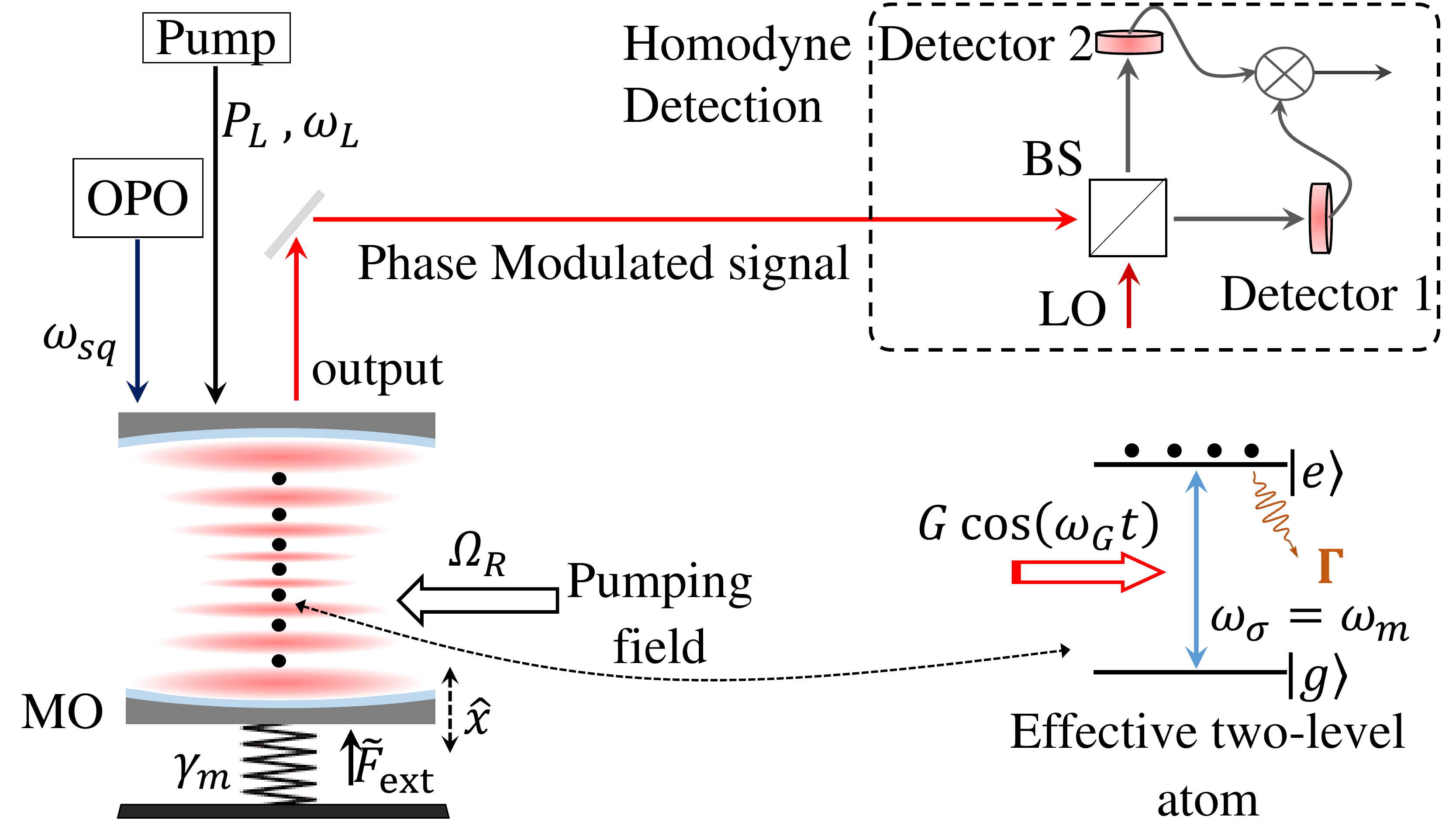}
	\end{center}
	\vspace*{-5mm}
	\caption{(Color online) Schematic description of the system under consideration. The system consists of a Fabry-P\'erot cavity, in which a single-mode MO is coupled
to the radiation pressure of the cavity field. Furthermore, the cavity contains an ensemble of effective two-level atoms with an effective transition rate
$\omega_{\sigma}=\omega_m$ that can be controlled by a classical laser field with Rabi frequency $\Omega_R$. The atomic ensemble behaves effectively as a negative-mass
oscillator under Faraday interaction and bosonization process \cite{motazedifard2016force}. An external classical force $\tilde{F}_{\rm ext}$ is exerted on the MO acting as
a sensor or test mass. The cavity is also driven by a coherent light field with power $P_L$ and frequency $\omega_L$. A squeezed vacuum light at resonance with the cavity
mode, $\omega_{\rm sq}=\omega_c$, is also injected into the cavity. The cavity output field enters the homodyne detection setup in order to extract information
on the external force and imprinted in the modulated phase of the cavity output field.}
	\label{fig1}
\end{figure}

As illustrated schematically in Fig.~\ref{fig1}, we consider a hybrid optomechanical system consisting of an optical mode of a Fabry-P\'erot cavity with resonance frequency
$\omega_c$ and a MO with an effective mass $m$, natural frequency $\omega_m$, and damping rate $\gamma_m$, coupled to the cavity field via radiation pressure
interaction and subjected to an external classical force $\tilde{F}_{\rm ext}$. Furthermore, the system contains an ensemble of $N_a$ effective two-level atoms trapped inside
the cavity and interacting with the cavity mode.
The cavity mode is coherently driven by a classical field of frequency $\omega_L$, input power $P_L$, and wavelength $\lambda_L$. Moreover, in the cavity it is injected into a
squeezed vacuum field, provided by the finite bandwidth output of an optical parametric oscillator (OPO), which is assumed to be resonant with the cavity mode.
The atomic ensemble interacts nonresonantly with the intracavity field and a classical control field. For
sufficiently large $N_a$, the atomic ensemble behaves effectively as a negative-mass oscillator (NMO) by assuming that the atoms are initially prepared in the higher-energy
level \cite{motazedifard2016force}.

After applying the bosonization procedure on the ultracold atomic ensemble using the Holstein-Primakoff transformation, the Hamiltonian of the system in the frame rotating at
driving laser frequency $\omega_L$ can be simplified as (for more details of the derivation of the Hamiltonian, see Ref.~\cite{motazedifard2016force})
\begin{align}
	\hat{H} =&\hbar\Delta_{c0}\hat{a}^\dagger\hat{a} + \hbar\omega_m\hat{b}^\dagger\hat{b}-\hbar\omega_m\hat{d}^\dagger\hat{d}+\hbar
g_0\hat{a}^\dagger\hat{a}(\hat{b}+\hat{b}^\dagger) \notag\\
	&  + \frac{\hbar}{2}G(\hat{a}+\hat{a}^\dagger)(\hat{d}+\hat{d}^\dagger) -i\hbar E_L(\hat{a}-\hat{a}^\dagger) + \hat{H}_{F},\label{eq1}
\end{align}
where $\Delta_{c0}=\omega_c-\omega_L$ is the cavity detuning, $E_L=\sqrt{P_L\kappa/\hbar\omega_L}$ is the pumping rate of the input laser with $\kappa$ the cavity
damping rate , $G$ denotes the collective atomic coupling with the cavity field, and $g_0=\omega_c x_{\rm zpf}/L$ is the single-photon optomechanical coupling strength, where
$x_{\rm zpf}=\sqrt{\hbar/2m\omega_m}$ is the zero-point fluctuation of the MO and $L$ is the resting length of the cavity.
Note that we have considered only the linear radiation pressure coupling between the MO and the cavity field. Furthermore, the operators $\hat{a}$, $\hat{b}$, and $\hat{d}$
are the annihilation operators of the cavity field, MO, and an effective negative-mass MO due to the bosonization of the atomic ensemble, respectively.
The first three terms of Eq.~\eqref{eq1} represent the free Hamiltonians of the cavity field, the MO, and the NMO, respectively. The fourth term of the Hamiltonian denotes
the optomechanical coupling between the cavity field and the MO, while the fifth term refers to the coupling between the atomic ensemble and the cavity field. The sixth term
accounts for the driving field and the last term stands for the contribution of the external classical force exerted on the MO, which is given by
\begin{equation}
	\hat{H}_F = \tilde{F}_{\rm ext}x_{\rm zpf}(\hat{b}+\hat{b}^\dagger).
\end{equation}

\section{\label{sec3}dynamics of the system}
The dynamics of the system is determined by the QLEs as
\begin{subequations}
	\begin{align}
\!\!\!	\dot{\hat{a}}&=\! -(i\Delta_{c0} \! +\! \frac{\kappa}{2})\hat{a}- \! ig_0\hat{a}(\hat{b} \!+ \! \hat{b}^\dagger) \! - \! i\frac{G}{2}(\hat{d} \!+ \! \hat{d}^\dagger)
\! + \! E_L \! \! +\! \! \! \sqrt{\kappa}\hat{a}^{\rm in} \! , \label{eq2a}\\
	\dot{\hat{b}}&=-(i\omega_m+\frac{\gamma_m}{2})\hat{b} -ig_0\hat{a}^\dagger\hat{a} +\frac{i}{\sqrt{2m\hbar\omega_m}}\big[\hat{\eta}(t)+\tilde{F}_{\rm ext}\big], \label{eq2b}\\
	\dot{\hat{d}}&=(i\omega_m-\frac{\Gamma}{2})\hat{d} -i\frac{G}{2}(\hat{a}+\hat{a}^\dagger)+\sqrt{\Gamma}\hat{d}^{\rm in}, \label{eq2c}
	\end{align}
	\label{eq2:all-lines}
\end{subequations}
where $\Gamma$ denotes the collective atomic dephasing rate. Here we have introduced three noise processes, which include the thermal noise acting on the MO that is
represented by the Brownian thermal noise operator $\hat{\eta}(t)$, the optical input vacuum noise $\hat{a}^{\rm in}$, and the optical vacuum fluctuations affecting the
atomic transitions represented by the bosonic operator $\hat{d}^{\rm in}$. These noises are uncorrelated, and the only nonvanishing correlation functions are
\begin{subequations}
	\begin{align}
		&\braket{\hat{a}^{\rm in}(t)\hat{a}^{\rm in}(t^\prime)^\dagger} = \braket{\hat{d}^{\rm in}(t)\hat{d}^{\rm in}(t^\prime)^\dagger}=\delta(t-t^\prime), \label{eq3a}\\
		&\braket{\hat{\eta}(t)\hat{\eta}(t^\prime)}\simeq\hbar m\gamma_m\big[\omega_m(2\bar{n}_m +1)\delta(t-t^\prime) + i\delta^\prime(t-t^\prime) \big], \label{eq3b}
	\end{align}
	\label{eq3:all-lines}
\end{subequations}
where $\bar{n}_m = \big[\exp(\hbar\omega_m/k_BT)-1\big]^{-1}$ is the mean phonon number of a thermal bath with temperature $T$ and $\delta^\prime(t-t^\prime)$ is the time
derivative of the Dirac delta function. In the derivation of Eq.~\eqref{eq3b}, we have assumed that the mechanical quality factor $Q_m=\omega_m/\gamma_m$ is very large,
justifying the weak damping limit where the Brownian noise can be treated as a Markovian noise \cite{giovannetti2001phase}.

We are interested in the regime where the cavity field and the atomic ensemble are strongly driven and the system is in the weak optomechanical coupling limit. Under these
conditions, we can linearize the dynamics of quantum fluctuations around the semiclassical steady state by considering system operators as
$\hat{A}\rightarrow\braket{\hat{A}}+\delta\hat{A}$ so that the higher orders of quantum fluctuation can be neglected with respect to the mean field $\braket{\hat{A}}$ in the
system dynamics. By introducing the amplitude and phase quadratures of the three modes of the system as
$\hat{X}_A=(\hat{A}+\hat{A}^\dagger)/\sqrt{2}$ and $\hat{P}_A=(\hat{A}-\hat{A}^\dagger)/i\sqrt{2}$, respectively, with $A= a,~b,~\mathrm{and}~d$, the linearized QLEs for the quantum
fluctuations are obtained as
\begin{subequations}
	\begin{align}
		\delta\dot{\hat{X}}_a &= \Delta_c\delta\hat{P}_a -\frac{\kappa}{2}\delta\hat{X}_a + \sqrt{\kappa}\hat{X}_a^{\rm in},\label{eq4a}\\
		\delta\dot{\hat{P}}_a&=-\Delta_c\delta\hat{X}_a - g\delta\hat{X}_b - G\delta\hat{X}_d - \frac{\kappa}{2}\delta\hat{P}_a + \sqrt{\kappa}\hat{P}_a^{\rm in},
\label{eq4b}\\
		\delta\dot{\hat{X}}_b &= \omega_m\delta\hat{P}_b, \label{eq4c}\\
		\delta\dot{\hat{P}}_b &= -\omega_m\delta\hat{X}_b - \gamma_m\delta\hat{P}_b - g\delta\hat{X}_a +\sqrt{\gamma_m}(\hat{f}+F_{\rm ext}),\label{eq4d}\\
		\delta\dot{\hat{X}}_d & = -\omega_m\delta\hat{P}_d -\frac{\Gamma}{2}\delta\hat{X}_d +\sqrt{\Gamma}\hat{X}_d^{\rm in}, \label{eq4e}\\
		\delta\dot{\hat{P}}_d & = \omega_m\delta\hat{X}_d - G\delta\hat{X}_a -\frac{\Gamma}{2}\delta\hat{P}_d + \sqrt{\Gamma}\hat{P}_d^{\rm in}, \label{eq4f}
	\end{align}
	\label{eq4:all-lines}
\end{subequations}
where $g=2g_0\alpha_s$ is the enhanced optomechanical coupling, $\Delta_c=\Delta_{c0} - 2g_0^2|\alpha_s|^2/\omega_m$ is the effective cavity detuning, and $\alpha_s$ is the
steady-state solution of the QLE for the cavity field which can always be considered as a real number without any loss of generality. We have also defined the scaled
mechanical thermal noise and external forces as $\hat{f}(t)=\hat{\eta}(t)/\sqrt{\hbar m\omega_m\gamma_m}$ and $F_{\rm ext}=\tilde{F}_{\rm ext}/\sqrt{\hbar
m\omega_m\gamma_m}$, respectively.

We mention that modeling the system dynamics with the linearized quantum Langevin equations is valid in the strong drive ($ n_{\rm cav} \gg 1$ or $ g/g_0 \gg 1 $) and weak single-photon optomechanical coupling regime ($ g_0 \ll \omega_m , \kappa $). 
Typically, with $10^3$ intracavity photons, linearization works already well and optomechanical force sensing is used just in this linearized regime.

In the next section, we will study the homodyne detection of a generic quadrature of the cavity output field aiming at force detection in the
presence of the CQNC condition and squeezed vacuum injection.

\section{\label{sec4} Enhancement of CQNC in the case of the homodyne measurement of a generic optical quadrature}
An external force acting on the MO shifts its position and changes the cavity effective length, which is directly reflected in the phase of the cavity output field.
Consequently, we have to measure the quadratures of the cavity output field to extract the signal associated with the external force. The quadrature of the cavity output
field relates to that of the intracavity field via the well-known input-output relation $\hat{a}^{\rm out}=\sqrt{\kappa}\delta\hat{a}-\hat{a}^{\rm in}$. In order to choose an
appropriate quadrature, we define the generalized quadrature of the cavity output field using a linear combination of both amplitude and phase as
\begin{equation} \label{eq5}
	\hat{P}_{a,\theta}^{\rm out}= \cos\theta\hat{P}_a^{\rm out}-\sin\theta\hat{X}_a^{\rm out},
\end{equation}
 where $\theta$ represents the phase angle of the local oscillator (LO) of the homodyne detector,
 which should be optimized so that the added noise imprinted to the output field spectrum is minimized.
 We recall that the optomechanical interaction imprints the optical amplitude quadrature noise on the mechanical resonator, which is then mapped onto the phase quadrature of the optical output field, a process which is also at the origin of ponderomotive squeezing. The resulting noise correlations between the optical amplitude and phase quadratures allow the measurement of unwanted noises to be reduced for an appropriate choice of homodyne phase angle and these correlations are those at the basis of the success of the variational readout method. Recently, variational homodyne readout has been employed in an ultracoherent optomechanical system, and off-resonant force and displacement sensitivity reaching $1.5~\mathrm{dB}$ below the SQL have been demonstrated \cite{Mason2019}.

 Now, by solving Eqs.~\eqref{eq4:all-lines} for $\delta\hat{X}_a$ and
 $\delta\hat{P}_a$ in the frequency domain and using the input-output relation, we obtain the generalized quadrature $\hat{P}_{a,\theta}^{\rm out}$ as
 \begin{align}
 	\hat{P}_{a,\theta}^{\rm out}(\omega)=&\sqrt{\kappa}\chi_a^\prime u_\theta\bigg\{-g\sqrt{\gamma_m}\chi_m(\hat{f}+F_{\rm ext})\notag \\
 	&+\sqrt{\kappa}\bigg[\bigg(1-\frac{1}{\kappa\chi_a^\prime}\bigg)\hat{P}_a^{\rm in}-\chi_a\Delta_c\hat{X}_a^{\rm in}  \bigg] \notag\\
 	&-G\sqrt{\Gamma}\chi_d \bigg[\hat{P}_d^{\rm in} -\frac{\Gamma/2 + i\omega}{\omega_m}\hat{X}_d^{\rm in} \bigg]\notag\\
 	&+\sqrt{\kappa}\chi_a(g^2\chi_m+G^2\chi_d)\hat{X}_a^{\rm in} \bigg\}\notag\\
 	&-\chi_a\Delta_c\sin\theta \hat{P}_a^{\rm in}-(\kappa\chi_a-1)\sin\theta\hat{X}_a^{\rm in}, \label{eq6}	
 \end{align}
in which we have defined $u_\theta=\cos\theta-\chi_a\Delta_c\sin\theta$, and also introduced the susceptibilities of the cavity field, the MO, and the atomic ensemble,
respectively as
 \begin{subequations}
 	\begin{align}
 		\chi_a(\omega)&=\frac{1}{\kappa/2+i\omega},\label{eq7a}\\
 		\chi_m(\omega)&=\frac{\omega_m}{(\omega_m^2-\omega^2)+i\omega\gamma_m},\label{eq7b}\\
 		\chi_d(\omega)&=\frac{-\omega_m}{(\omega_m^2-\omega^2+\Gamma^2/4)+i\omega\Gamma},\label{eq7c}
 	\end{align}
 	\label{eq7:all-lines}
 \end{subequations}
and the modified cavity mode susceptibility as
\begin{equation}\label{eq8}
	\frac{1}{\chi_a^\prime(\omega)} = \frac{1}{\chi_a(\omega)} -\chi_a(\omega)\Delta_c\big[g^2\chi_m(\omega)+G^2\chi_d(\omega)-\Delta_c\big].
\end{equation}
Equation~\eqref{eq6} denotes the experimental signal which has to be measured to estimate the external force $F_{\rm ext}$. We define the generalized estimated external force as
 \begin{align}
 	\hat{F}_{\rm ext,\theta}^{\rm est}\equiv\frac{-1}{g\sqrt{\kappa\gamma_m}\chi_a^\prime\chi_m u_\theta}\hat{P}_{a,\theta}^{\rm out}\equiv F_{\rm ext}+\hat{F}_{N,\theta},
 \label{eq9}
 \end{align}
 in which $\hat{F}_{N,\theta}$ is the generalized added force noise, given by
 \begin{align}
 	\hat{F}_{N,\theta}=&\hat{f}-\sqrt{\frac{\kappa}{\gamma_m}}\frac{1}{g\chi_m}\bigg[\bigg(1-\frac{1}{\kappa\chi_a^\prime}\bigg)\hat{P}_a^{\rm
 in}-\chi_a\Delta_c\hat{X}_a^{\rm in} \bigg]\notag\\
 	&+\sqrt{\frac{\Gamma}{\gamma_m}}\frac{G\chi_d}{g\chi_m}\bigg[\hat{P}_d^{\rm in}-\frac{\Gamma/2+i\omega}{\omega_m}\hat{X}_d^{\rm in} \bigg]\notag\\
 	&-\sqrt{\frac{\kappa}{\gamma_m}}\frac{g^2\chi_m+G^2\chi_d}{g\chi_m}\chi_a\hat{X}_a^{\rm in}\notag\\
 	&+\frac{\mathcal{B}}{g\sqrt{\kappa\gamma_m}\chi_a^\prime\chi_m}\big[\chi_a\Delta_c\hat{P}_a^{\rm in}+\big(\kappa\chi_a-1\big)\hat{X}_a^{\rm in}\big]. \label{eq10}
 \end{align}
Here, we have defined $\mathcal{B}=\sin\theta/u_{\theta}$. According to Eq.~\eqref{eq10}, there are five different contributions to the added force noise. The first term
corresponds to the thermal noise of the MO. The second term refers to the shot noise contribution associated with the cavity field, which is significant especially at low
driving power, and it is modified by the squeezed injection, as demonstrated in \cite{motazedifard2016force}. The third term represents the atomic noise originating
from the interaction between the atomic ensemble and the cavity field. The fourth term denotes the backaction noise due to the radiation pressure coupling of the cavity field
with the MO and the atomic ensemble, which grows by increasing the strength of the cavity field. The last term represents the contribution to the added force noise
arising from the phase of the homodyne detection of the cavity output field. Only a moment's thought is needed to conclude that this term vanishes for $\theta=0$.

\subsection{\label{subsec4A}Generalized force noise power spectral density}
To quantify the sensitivity of the force measurement, we define the generalized force noise power spectral density as
\begin{equation}\label{eq11}
	S_{F,\theta}(\omega)\delta(\omega-\omega^\prime)=\frac{1}{2}\big[\braket{\hat{F}_{N,\theta}(\omega)\hat{F}_{N,\theta}(-\omega^\prime)}+\mathrm{c.c.} \big].
\end{equation}
In the steady state, $\kappa\gg\omega$, and in the presence of the squeezed vacuum injection, the generalized spectral density of the added force noise is given by (see
Appendix \ref{App.A})
\begin{align}
	S_{F,\theta}(\omega)=&S_{th}(\omega) + S_{f}(\omega) + S_{at}(\omega) + S_b(\omega) + S_h(\omega)\notag\\
	&+S_{fb}(\omega) + S_{fh}(\omega) + S_{bh}(\omega),\label{eq12}
\end{align}
where the first five terms correspond to the noise contributions of the Brownian motion of the MO, the cavity field, the atomic ensemble, the BA, and the homodyne phase,
respectively:
\begin{equation} \label{eq13}
	S_{th}(\omega)=k_BT/\hbar\omega_m,
\end{equation}
\begin{align}
	S_f(\omega)=&\frac{\kappa}{g^2\gamma_m|\chi_m(\omega)|^2}\Bigg\{\Delta_c\mathrm{Im}\big[Z(\omega)(1-2i\mathrm{Im}M)\big] \notag\\ &+\Bigg(1+\frac{1}{\kappa^2|\chi_a^\prime(\omega)|^2}-\frac{2\mathrm{Re}\chi_a^\prime(\omega)}{\kappa|\chi_a^\prime(\omega)|^2}\Bigg)\Big(N+\frac{1}{2}-\mathrm{Re}M\Big)\notag\\
	&+\frac{4\Delta_c^2}{\kappa^2}\Big(N+\frac{1}{2}+\mathrm{Re}M\Big) \Bigg\}, \label{eq14}
\end{align}
\begin{equation}\label{eq15}
	S_{at}(\omega)=\frac{|\mathcal{A}(\omega)|^2}{2}\bigg(1+\frac{\omega^2+\Gamma^2/4}{\omega_m^2}\bigg),
\end{equation}
\begin{equation}\label{eq16}
	S_b(\omega)=\frac{4 g^2}{\kappa\gamma_m}\bigg|1+\frac{G^2}{g^2}R(\omega)\bigg|^2\Big(N+\frac{1}{2}+\mathrm{Re}M\Big),
\end{equation}
\begin{align} S_h(\omega)=&\frac{2\mathcal{B}^2}{g^2\kappa\gamma_m|\chi_a^\prime(\omega)|^2|\chi_m(\omega)|^2}\Bigg\{\bigg(\frac{1}{2}+\frac{2\Delta_c^2}{\kappa^2}\bigg)\bigg(N+\frac{1}{2}\bigg)\notag\\
	&+\bigg(\frac{1}{2}-\frac{2\Delta_c^2}{\kappa^2}\bigg)\mathrm{Re}M+\frac{2\Delta_c}{\kappa}\mathrm{Im}M \Bigg\}.\label{eq17}
\end{align}
Furthermore, the last three terms of Eq.~\eqref{eq12} refer to the quantum interferences associated with the joint action of the different modes of the system. The term
$S_{fb}(\omega)$ relates to the quantum interference between the cavity field and the atomic ensemble, $S_{fh}(\omega)$ originates from the quantum interference between the
cavity field and the homodyne phase contribution, and $S_{bh}(\omega)$ corresponds to the backaction-homodyne phase interference:
\begin{align}
	S_{fb}(\omega)=&\frac{\kappa}{\gamma_m}\mathrm{Im}\Bigg[(2i\mathrm{Im}M-1)\frac{Z(\omega)}{\chi_m(-\omega)}\Bigg(1+\frac{G^2}{g^2}R(\omega)\Bigg) \Bigg]\notag\\
	&-\frac{8\Delta_c}{\kappa\gamma_m}\mathrm{Re}\Bigg[\frac{1+(G^2/g^2)R(\omega)}{\chi_m(-\omega)}\Bigg]\bigg(N+\frac{1}{2}+\mathrm{Re}M
	\bigg),\label{eq18}
\end{align}
\begin{align} S_{fh}(\omega)=&-\frac{\mathcal{B}}{g^2\gamma_m|\chi_m(\omega)|^2}\Bigg\{2\Delta_c\mathrm{Re}\bigg[\frac{Z(\omega)}{\chi_a^\prime(-\omega)}\bigg]\bigg(N+\frac{1}{2}-\mathrm{Re}M\bigg)\notag\\
	&-2\Delta_c\mathrm{Re}\bigg[\frac{\chi_a(\omega)}{\chi_a^\prime(-\omega)}\bigg]\bigg(N+\frac{1}{2}+\mathrm{Re}M\bigg)\notag\\
	&-\mathrm{Im}\Bigg[(2i\mathrm{Im}M+1)\bigg(\frac{4\Delta_c^2/\kappa^2}{\chi_a^\prime(\omega)}+\frac{1-\kappa Z(-\omega)}{\chi_a^\prime(-\omega)}\notag\\
	& -\frac{1}{\kappa|\chi_a^\prime(\omega)|^2} \bigg) \Bigg] \Bigg\},\label{eq19}
\end{align}
\begin{align}
S_{bh}(\omega)=&-\frac{2\mathcal{B}}{\kappa\gamma_m}\Bigg\{\frac{2\Delta_c}{\kappa}\mathrm{Im}\Bigg[(2i\mathrm{Im}M+1)\frac{1+(G^2/g^2)R(\omega)}{\chi_m(\omega)\chi_a^\prime(\omega)}\Bigg]\notag\\
	&+2\mathrm{Re}\Bigg[\frac{1+(G^2/g^2)R(\omega)}{\chi_m(\omega)\chi_a^\prime(\omega)} \Bigg]\bigg(N+\frac{1}{2}+\mathrm{Re}M\bigg) \Bigg\},
\end{align}
where
\begin{subequations}
	\begin{align}
	Z(\omega)&=\chi_a(\omega)\bigg(1-\frac{1}{\kappa\chi_a^\prime(-\omega)}\bigg),\label{eq21a}\\
	R(\omega)&=\frac{\chi_d(\omega)}{\chi_m(\omega)},\label{eq21b}\\
	\mathcal{A}(\omega)&=\frac{G}{g}\sqrt{\frac{\Gamma}{\gamma_m}}R(\omega). \label{eq21c}
	\end{align}
\end{subequations}
Note that in the derivation of the power spectral density given by Eq.~\eqref{eq12}, we used the fact that in the limit of $\omega/\kappa \ll 1$ (resolved-sideband regime), the optical susceptibility can be approximated as  $\chi_a\simeq2/\kappa$ (see Appendix \ref{app. C}). Moreover, the squeezing parameters $M$ and $N$ relate to the effective second-order nonlinearity $\epsilon$ and cavity decay rate $\gamma$ of the OPO as $M=(\epsilon\gamma/2)(1/b_x^2 + 1/b_y^2)$ and $N=(|\epsilon|\gamma/2)(1/b_x^2 - 1/b_y^2)$ with $b_x=\gamma/2-|\epsilon|$ and $b_y=\gamma/2+|\epsilon|$ (see Appendix \ref{app.D} for experimental constraints on the squeezing parameter $ N $ to consider its maximum value in the following graphs in figures). We recall that with the chosen units, the noise spectral density is dimensionless and in order to convert it to
$\mathrm{N}^2~\mathrm{Hz}^{-1}$ units we have to multiply by the scale factor $\hbar m \omega_m \gamma_m$.

\subsection{CQNC Conditions\label{SubSecA}}
The CQNC effect, which refers to the perfect cancellation of the backaction noise at all frequencies, leads to significant noise suppression in force detection. According to
the fourth term of Eq.~\eqref{eq10}, for $g=G$ and $\chi_d=-\chi_m$, the mechanical backaction and the atomic backaction contributions to the added force noise cancel each
other at all frequencies and hence they offer noise and antinoise paths to the signal force. Strictly speaking, the CQNC refers to the perfect matching of 
(i) the mirror-field with the atom-field couplings, i.e., $g=G$, 
(ii) the mechanical dissipation rate with the dephasing rate of the atoms, i.e., $\Gamma=\gamma_m$, and
(iii) the mechanical susceptibility $\chi_m$ with the atomic susceptibility $\chi_d$, which is realized when the MO has a high quality factor.

One can easily conclude that in the presence of the  perfect CQNC conditions we have $1+(G^2/g^2)R(\omega)=0$ and hence, $S_b(\omega)=S_{fb}(\omega)=S_{bh}(\omega)=0$. As a consequence, under perfect CQNC conditions, the generalized spectral density of the added force given by Eq.~\eqref{eq12} reduces to
\begin{align}
S_{F,\theta}(\omega)=&\frac{k_BT}{\hbar\omega_m}+\frac{1}{2}\bigg(1+\frac{\omega^2+\gamma_m^2/4}{\omega_m^2}\bigg)\notag\\
&+\frac{\kappa}{g^2\gamma_m|\chi_m(\omega)|^2}\bigg[\frac{1}{2}\mu_{+}^2+\Sigma(N,M,y)+\Theta(N,M,y,\theta) \bigg], \label{eq22}
\end{align}
where we have introduced the normalized detuning as $y\equiv\Delta_c/\kappa$ and $\mu_{\pm}\equiv\frac{1}{2}\pm2y^2$. The functions $\Sigma$ and $\Theta$ represent the contributions
of the injected squeezing and the homodyne phase to the optomechanical shot noise, respectively, and are given by
\begin{subequations}
	\begin{align}
	\Sigma(N,M,y)&=\mu_{+}^2N + (8y^2 - \mu_{+}^2)\mathrm{Re}M - 4y\mu_{-}\mathrm{Im}M,\label{eq23a}\\
	\Theta(N,M,y,\theta)&=2\mathcal{B}^2\bigg[\mu_{+}^3\bigg(N+\frac{1}{2}\bigg) + \mu_{-}\mu_{+}^2\mathrm{Re}M + 2y\mu_{+}^2\mathrm{Im}M \bigg]\notag\\
	&\quad +\mathcal{B}\bigg[4y\mu_{+}^2\bigg(N+\frac{1}{2}\bigg) + 4y\mu_{+}(\mu_{-}+1)\mathrm{Re}M \notag\\
	&\quad-2\mu_{+}(\mu_{-}-4y^2)\mathrm{Im}M \bigg].\label{eq23b}
	\end{align}
\end{subequations}
Note that with the chosen units, the spectral density of the added force noise is dimensionless, and it should be multiplied by the scale factor $\hbar m\omega_m\gamma_m$ in order to convert it to $\mathrm{N}^2~\mathrm{Hz}^{-1}$ units. According to Eq.~\eqref{eq22}, the homodyne phase contribution to the noise spectrum can be considered as a shot-noise-like term and consequently we expect that it decreases by increasing the strength of the cavity field.

We have to compare the noise spectrum in our scheme with that of a bare optomechanical setup, which serves as a standard system formed by an optical cavity coupled to a MO.
The SQL for stationary force detection comes from the minimization of the force spectrum of the standard system at a given frequency over the driving power, which is given by
\cite{motazedifard2016force}
\begin{equation} \label{eq24}
S_{\rm SQL}(\omega)=\frac{1}{\gamma_m|\chi_m(\omega)|}.
\end{equation}

As mentioned in \cite{motazedifard2016force}, by considering $M=|M|\exp(i\phi)$, in the absence of the homodyne detection ($\theta=0$), the optimized parameters for the
injected squeezing in order to suppress the shot noise contributions as much as possible are $|M|=\sqrt{N(N+1)}$ (pure squeezing) and
\begin{equation} \label{eq25}
\phi_{\rm opt}(y)=\frac{4y\mu_{-}}{\mu_{+}^2-8y^2},
\end{equation}
which can be obtained by minimizing the noise spectrum given by Eq.~\eqref{eq22} over the phase of the squeezing parameter $\phi$ for $\theta=0$. Since we want to make a
proper comparison between the cases of $\theta=0$ and $\theta\ne0$, we have assumed that the phase of the squeezing parameter in our scheme, even for $\theta\ne 0$,
is given by Eq.~\eqref{eq25}.

The appropriate quadrature of the cavity output field is determined by minimizing the noise spectrum over the homodyne phase angle $\theta$, which under the perfect CQNC
conditions, yields
\begin{equation}\label{eq26}
	\tan\theta_{\rm opt} = -\frac{K/2L}{1-y(K/L)},
\end{equation}
where
\begin{subequations}
	\begin{align}
	K&= 4y\mu_{+}^2\bigg(N+\frac{1}{2}\bigg) + 4y\mu_{+}(\mu_{-}+1)\mathrm{Re}M -2\mu_{+}(\mu_{-}-4y^2)\mathrm{Im}M, \label{eq26a}\\
	L&=2\mu_{+}^3\bigg(N+\frac{1}{2}\bigg) + 2\mu_{-}\mu_{+}^2\mathrm{Re}M + 4y\mu_{+}^2\mathrm{Im}M. \label{eq26b}
	\end{align}
	\label{eq27:all-lines}
\end{subequations}
By substituting Eqs.~\eqref{eq27:all-lines} in Eq.~\eqref{eq22}, the minimized noise spectrum in the presence of the perfect CQNC is obtained as
\begin{align}
S_{F,\theta_{\rm opt}}^{\rm min}(\omega)=&\frac{k_BT}{\hbar\omega_m} + \frac{1}{2}\bigg(1+\frac{\omega^2+\gamma_m^2/4}{\omega_m^2}\bigg) \notag\\
&+ \frac{\kappa}{g^2\gamma_m|\chi_m(\omega)|^2}\bigg[\frac{1}{2}\mu_{+}^2 + \Sigma(N,M,y) - \frac{K^2}{4L} \bigg]. \label{eq28}
\end{align}
In the case when the cavity is driven at resonance, i.e., $y=0$, one can easily conclude that $\phi_{\rm opt}=0$, $K=0$, and hence, $\theta_{\rm opt}=0$. Consequently, in this case, the homodyne phase contribution to the shot noise vanishes and does not improve the sensitivity of the force detection. This means that under perfect CQNC conditions and
at cavity resonance ($\Delta_c=0$), the phase quadrature $\hat{P}_a^{\rm out}$ is the optimal output quadrature which has to be measured to extract the
external force signal.
In the case of off-resonance cavity driving, i.e., $y\ne0$, as demonstrated in Eq.~\eqref{eq28}, in the regime where the function $L$ given by Eq.~\eqref{eq27:all-lines}
becomes positive, the homodyne phase contribution to the shot noise leads to the reduction of the noise spectrum, which results in the improvement of the force detection
sensitivity.
Note that the advantage of our scheme against the simple phase detection scheme in Ref.~\cite{motazedifard2016force} is given by
\begin{align} \label{eqold30}
	\delta S_{\rm CQNC} &\equiv S_{F,\theta_{\rm opt}}^{\rm min}(\omega) - S_{F,\theta=0}(\omega)\notag\\
	& = -\frac{\kappa}{4g^2\gamma_m|\chi_m(\omega)|^2}\bigg(\frac{K^2}{L}\bigg),
\end{align}
where $\delta S_{\rm CQNC}$ refers to a noise reduction advantage, which can be redefined in decibel scale as
\begin{align} 
	\delta S_{\rm CQNC} =&10 \log\Bigg(\frac{S_{F,\theta_{\rm opt}}^{\rm min}(\omega)}{S_{F,\theta=0}(\omega)} \Bigg). \label{eq.31_0}
\end{align}
\begin{figure}
	\begin{center}
		\includegraphics[width=8cm]{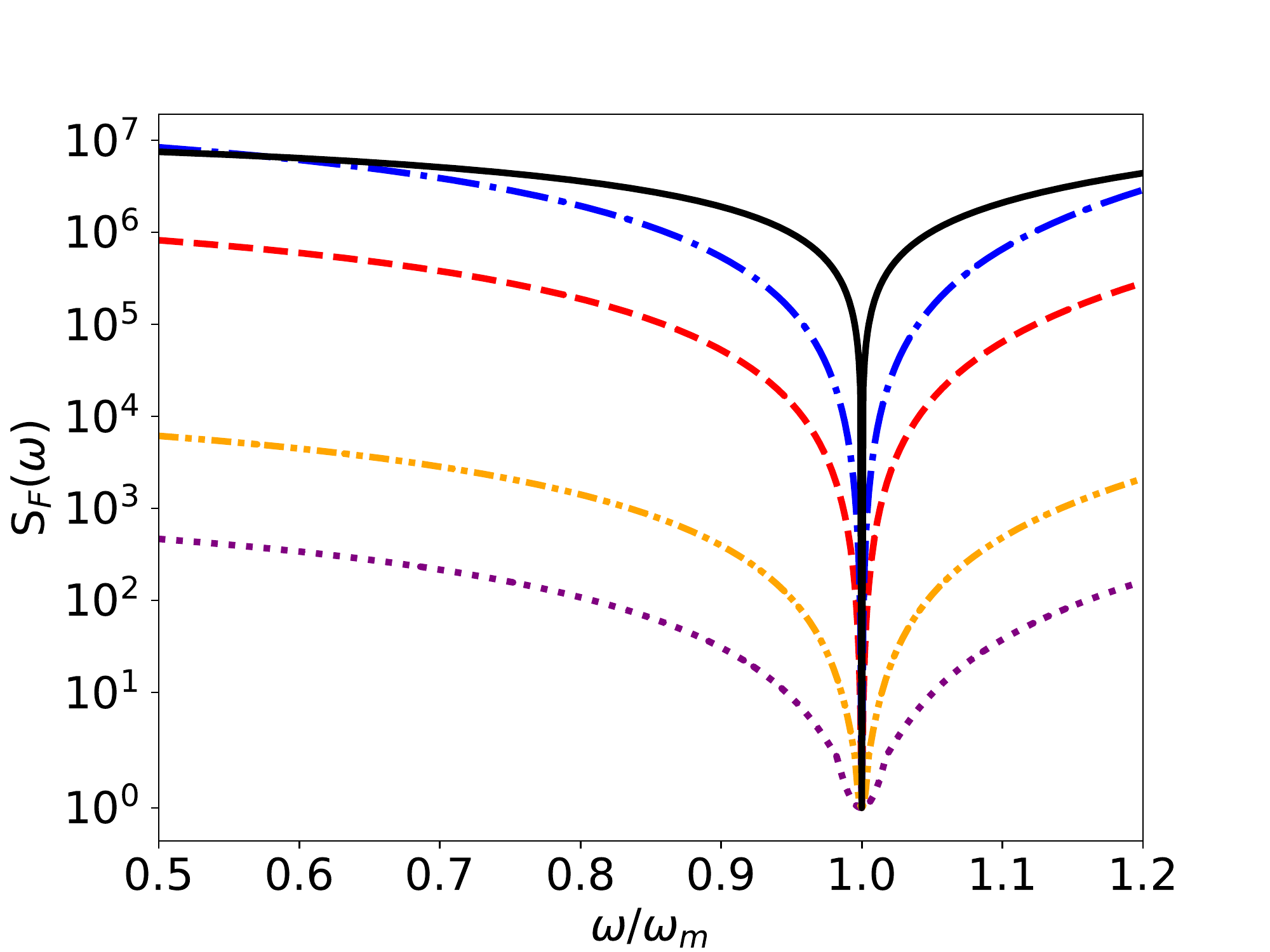}
	\end{center}
	\vspace*{-5mm}
	\caption{(Color online) Force noise spectral density versus $\omega/\omega_m$, with an optimized squeezed injected light, i.e., $|M|=\sqrt{N(N+1)}$ with $N=10$, and
$\phi=\phi_{\rm opt}$. The results are plotted for the case of off-resonance cavity driving, with $y=1/2$ (red dashed line for $\theta=0$ and purple dotted line for
$\theta=\theta_{\rm opt}$) and $y=1$ (blue dash-dotted line for $\theta=0$ and orange dash--double-dotted line for $\theta=\theta_{\rm opt}$). The black solid line corresponds
to the SQL. The other parameters are $\omega_m/2\pi=300~\mathrm{kHz}$, $\gamma_m/2\pi=30~\mathrm{mHz}$, $g_0/2\pi=300~\mathrm{Hz}$, $\lambda_L=780~\mathrm{nm}$,
$P_L=24~\mu\mathrm{W}$ (i.e., $g/g_0=4.91\times10^3$), and $\kappa/2\pi=10~\mathrm{MHz}$. }
	\label{fig2}
\end{figure}
\begin{figure}
	\begin{center}
		\includegraphics[width=8cm]{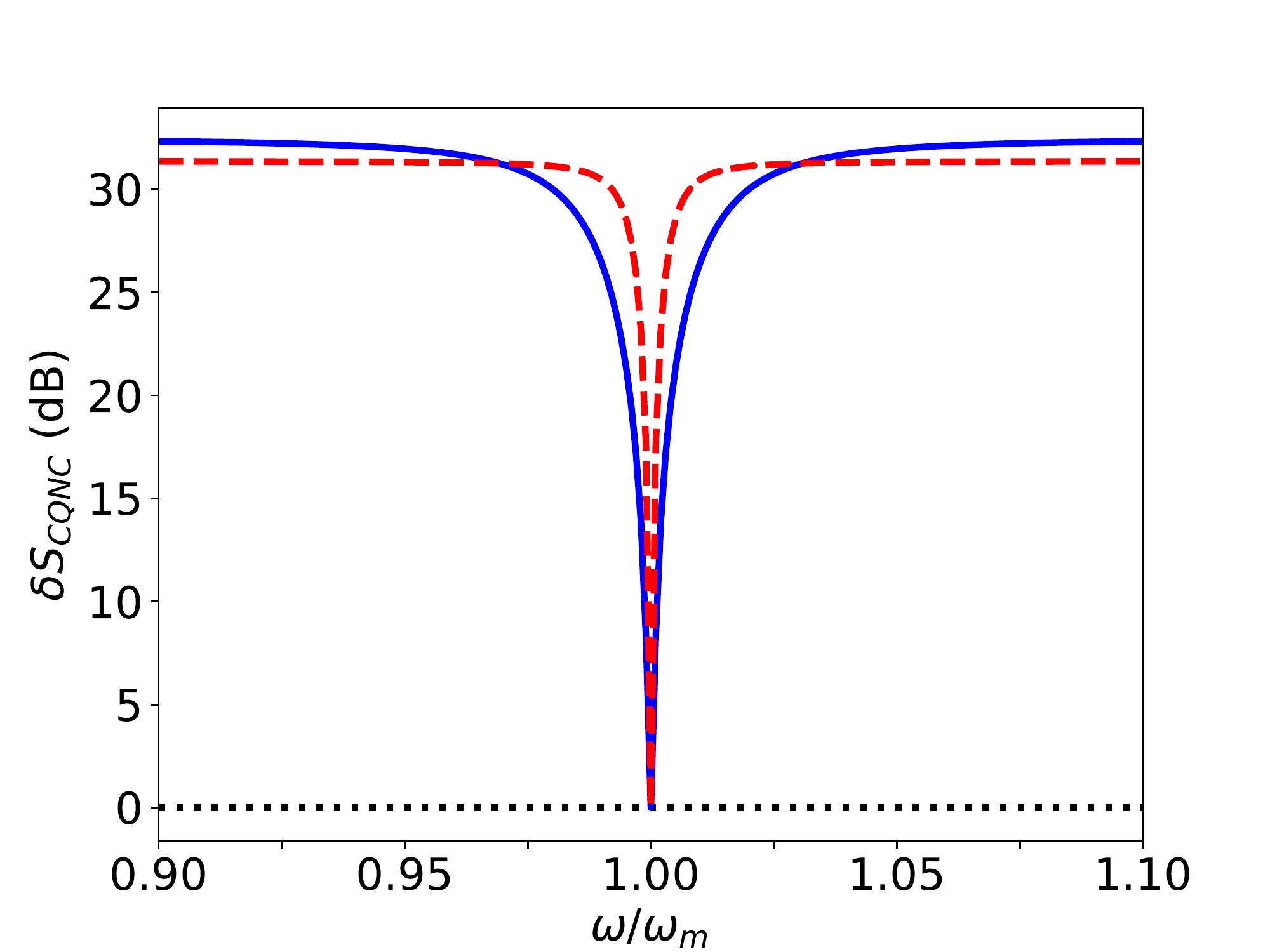}
	\end{center}
	\vspace*{-5mm}
	\caption{(Color online) Noise reduction advantage $\delta S_{\rm ICQNC}$ in decibel scale versus $\omega/\omega_m$, with an optimized squeezed injected light, i.e., $|M|=\sqrt{N(N+1)}$ with $N=10$, and
		$\phi=\phi_{\rm opt}$. Different curves correspond to  $y=0$ (back dotted line), $y=0.5$ (blue solid  line), and $y=1$ (red dashed line). The other parameter values are the same as those in Fig.~\ref{fig2}.  }
	\label{fig2_1}
\end{figure}
\begin{figure}
	\begin{center}
		\includegraphics[width=8cm]{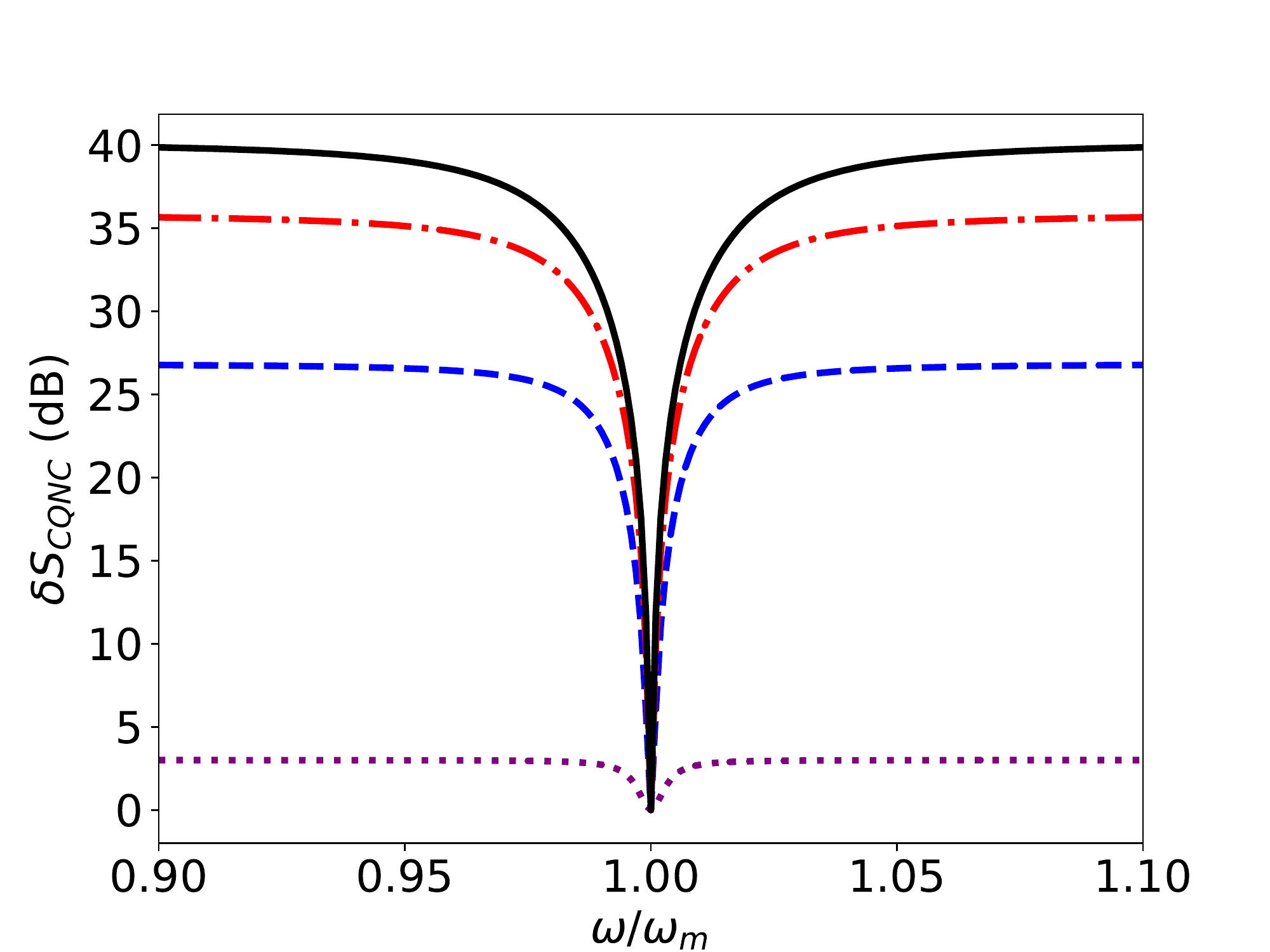}
	\end{center}
	\vspace*{-5mm}
	\caption{(Color online) Noise reduction advantage in decibel scale versus $\omega/\omega_m$. Here we set the normalized detuning $y=\frac{1}{2}$ and consider different values for the squeezing parameter: $N=0$ (purple dotted  line), $N=5$ (blue dashed  line), $N=15$ (red dash-dotted  line), and $N=25$ (black solid line). The other parameter values are the same as those in Fig.~\ref{fig2}. }
	\label{fig3}
\end{figure}
\begin{figure}
	\begin{center}
		\includegraphics[width=8cm]{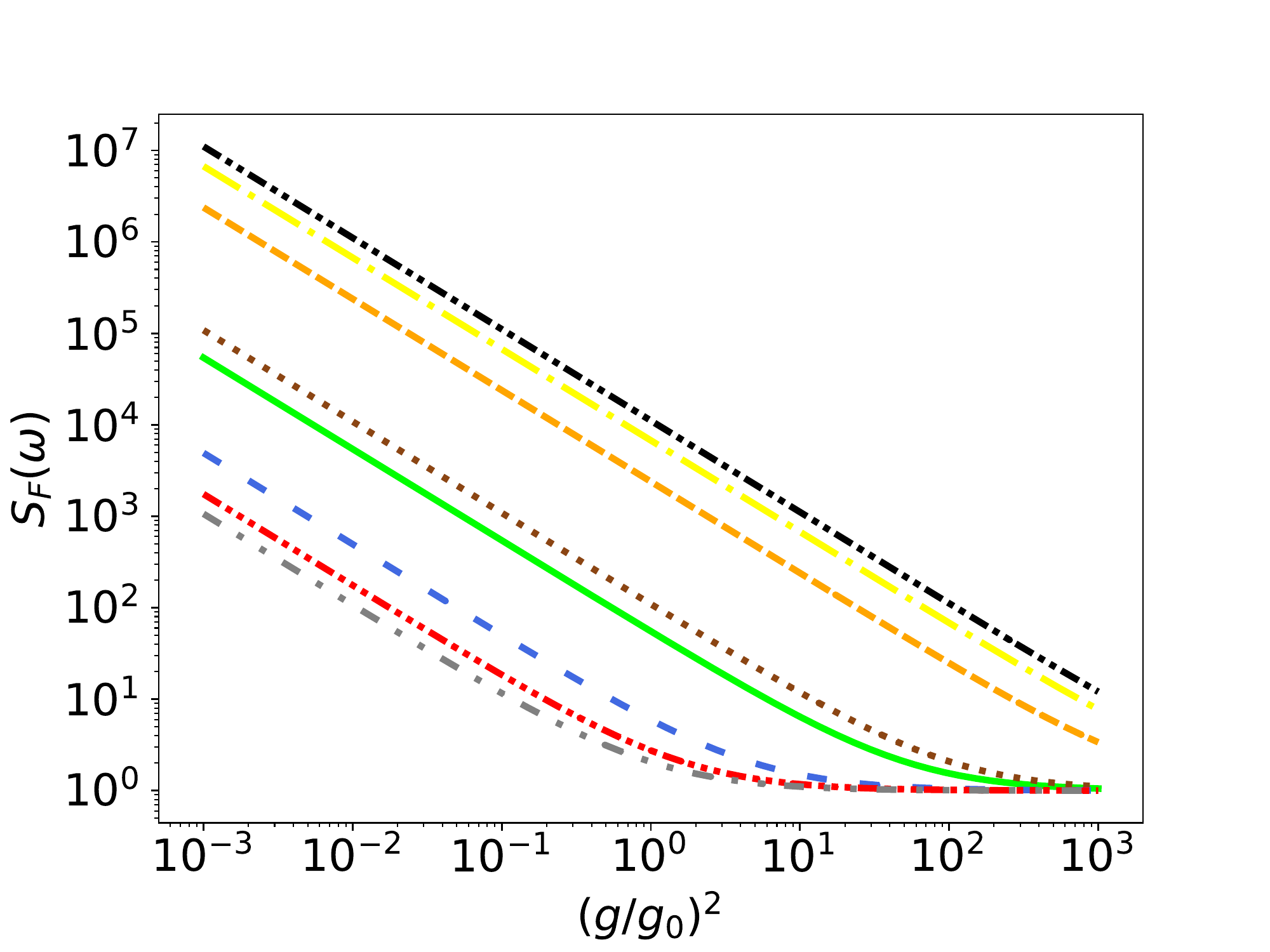}
	\end{center}
	\vspace*{-5mm}
	\caption{(Color online) Force noise spectral density versus $(g/g_0)^2$, at the frequency $\omega=\omega_m+4\gamma_m$. Here we set the normalized detuning $y=\frac{1}{2}$, and consider different values for the squeezing parameter:
	$N=0$ (green solid line for $\theta=\theta_{\rm opt}$ and brown dotted line for $\theta=0$),
	$N=5$ (blue dashed line for $\theta=\theta_{\rm opt}$ and orange densely dashed line for $\theta=0$),
	$N=15$ (red dash--triple-dotted line for $\theta=\theta_{\rm opt}$ and yellow dash-dotted line for $\theta=0$), and
	$N=25$ (gray dash--double-dotted line for $\theta=\theta_{\rm opt}$ and black  densely dash--double-dotted line for $\theta=0$). The other parameter values are the same as those in Fig.~\ref{fig2}. }
	\label{fig4}
\end{figure}
\begin{figure}
	\begin{center}
		\includegraphics[width=8cm]{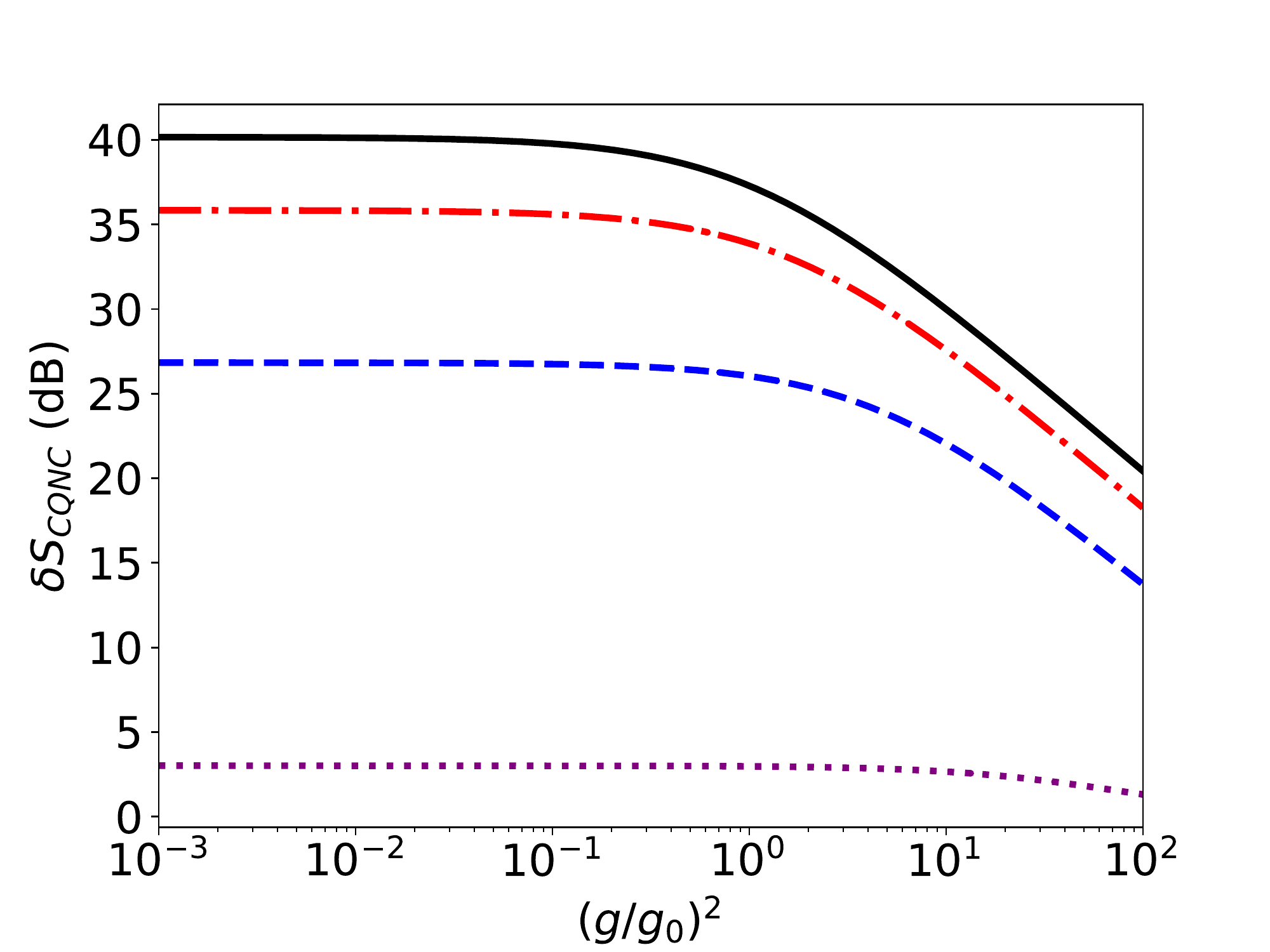}
	\end{center}
	\vspace*{-5mm}
	\caption{(Color online) Noise reduction advantage in decibel scale versus $(g/g_0)^2$, at the frequency $\omega=\omega_m+4\gamma_m$. Here we set the normalized detuning $y=\frac{1}{2}$, and consider different values for the squeezing parameter: $N=0$ (purple dotted line), $N=5$ (blue dashed line), $N=15$ (red dash-dotted line), and $N=25$ (black solid line). The other parameter values are the same as those in Fig.~\ref{fig2}. }
	\label{fig5}
\end{figure}

Let us now illustrate how the contribution of the rotated quadrature defined by Eq.~\eqref{eq5}, which is equivalent to the effect of the homodyne phase, together with the squeezed vacuum field injection into the cavity under the CQNC conditions, affects the noise cancellation. 
We consider the experimentally feasible system parameters
\cite{motazedifard2016force} $\omega_m/2\pi=300~\mathrm{kHz}$, $\gamma_m/2\pi=30~\mathrm{mHz}$, $g_0/2\pi=300~\mathrm{Hz}$, $\lambda_L=780~\mathrm{nm}$, $P_L=24~\mu\mathrm{W}$ (i.e., $g/g_0=2\alpha_s=2\sqrt{\bar n_{\rm cav}}=2 \sqrt{P_L/\hbar\omega_L \kappa}=4.91\times10^3$), $\kappa/2\pi=10\mathrm{MHz}$, and effective mass $m=1~\rm ng$.

Since under perfect CQNC conditions and resonant cavity driving ($\Delta=0$) the optimized quadrature is the phase quadrature, which has been investigated in
Ref.~\cite{motazedifard2016force}, in Sec.~\ref{SubSecA} we focus our attention on the nonzero detuning case in the perfect CQNC condition. In Sec.~\ref{sub4c} we investigate both zero and nonzero detuning in the imperfect CQNC condition.

In Fig.~\ref{fig2} the spectral density of the force noise optimized over the squeezing parameters [$M=\sqrt{N(N+1)}$ and $\phi_{\rm opt}$] is plotted versus the normalized
frequency $\omega/\omega_m$. This plot clearly shows the advantage of the homodyne detection with an optimized phase angle $\theta_{\rm opt}$ in reducing the noises
present in the force spectrum. 
In Fig.~\ref{fig2_1} we plot the spectrum of the noise reduction advantage $\delta S_{\rm ICQNC}$ given by Eq.~\eqref{eq.31_0} for different detunings. This illustrates that in the close vicinity of mechanical resonance frequency $\omega_m$, the noise reduction advantage is greater for larger values of cavity detuning, while it vanishes when the cavity drives resonantly, as expected.  
The spectrum of the noise reduction advantage $\delta S_{\mathrm{CQNC}}$ for different values of squeezing parameter $N$  is plotted in Fig.~\ref{fig3}. This figure reveals that the noise reduction advantage
increases at all frequencies as the squeezing parameter $N$ grows.
The effect of the cavity driving power, which is proportional to $(g/g_0)^2$, on the force noise
spectrum is illustrated in Fig.~\ref{fig4}. 
This figure shows that the variational homodyne CQNC reduces the force noise spectrum, notably at low driving powers. In fact, homodyne detection at the optimized phase angle $\theta_{\rm opt}$ leads to the third term in the second line of Eq.~(\ref{eq28}), which is shot-noise-like, i.e., proportional to $1/g^2$, and negative, hence reducing the total shot-noise contribution to the force noise spectrum compared to the standard case when $ \theta=0 $ [see Eqs.~(\ref{eq28}) and (\ref{eqold30})].

This result can also be observed in Fig.~\ref{fig5}, which illustrates the effect of the cavity driving power on the noise reduction advantage. As it is evident, the noise reduction advantage diminishes as the cavity driving power increases. Moreover, increasing the squeezing parameter $N$ leads to an increase of the noise reduction advantage, which, interestingly, rises up to $40~\mathrm{dB}$ for $N=25$.

\subsection{Imperfect CQNC Conditions \label{sub4c}}
\subsubsection{Resonant case $(\Delta_c=0)$}
Perfect backaction noise cancellation requires that the mechanical parameters match perfectly to the atomic ones. As discussed in \cite{motazedifard2016force}, one can
make the resonance frequency of the collective atomic mode equal to the mechanical resonance frequency $\omega_m$ by tuning the magnetic field exerted on the atomic
ensemble. Moreover, by adjusting the cavity and atomic driving rates, one can match the
field-atom coupling $G$ to the field-mirror coupling $g$, and also match the two decay rates $\Gamma$ and $\gamma_m$. However, in practice, matching the two latter parameters is more involved and therefore it is relevant to investigate if variational homodyne detection can be useful in the absence of perfect matching of the coupling and decay rates.
First, we consider the zero detuning case, i.e., $\Delta_c=0$. Here we still consider the optimum values for the squeezing parameters obtained earlier under the CQNC
conditions, i.e., optimized pure squeezing with $|M|=\sqrt{N(N+1)}$ and $\phi=\phi_{\rm opt}(0)=0$. Minimizing the generalized spectral density of the force noise given by
Eq.~\eqref{eq12} over the homodyne phase angle $\theta$ yields
\begin{equation}\label{eq30}
\tan\theta_{\rm opt}=\frac{4g^2}{\kappa}|\chi_m(\omega)|^2\mathrm{Re}\Bigg[\frac{1+(G^2/g^2)R(\omega)}{\chi_m(\omega)}\Bigg]
\end{equation}
for the optimum homodyne phase.
By substituting Eq.~\eqref{eq30} in Eq.~\eqref{eq12}, one can obtain the optimized force noise spectrum as
\begin{align}
&S_{F,\theta_{\rm opt}}^{\rm min}(\omega)=\frac{k_BT}{\hbar\omega_m} + \frac{\Gamma}{2\gamma_m}\frac{G^2}{g^2}R(\omega)\Bigg(1+\frac{\omega^2+\Gamma^2/4}{\omega_m^2}
\Bigg)\notag\\
&\quad + \frac{\kappa}{g^2\gamma_m|\chi_m(\omega)|^2}\bigg(N+\frac{1}{2}-\sqrt{N(N+1)}\bigg)\notag\\
&\quad +\frac{4g^2}{\kappa\gamma_m}\bigg|1+\frac{G^2}{g^2}R(\omega) \bigg|^2 \bigg(N+\frac{1}{2}+\sqrt{N(N+1)}\bigg)\notag\\
&\quad -\frac{4g^2}{\kappa\gamma_m}\Bigg[\mathrm{Re}\big(\frac{1+(G^2/g^2)R(\omega)}{\chi_m(\omega)}\big)\Bigg]^2\frac{|\chi_m(\omega)|^2}{N+\frac{1}{2}+\sqrt{N(N+1)}}.
\label{eq31}
\end{align}
Here the last term corresponds to the subtraction associated with the choice of the optimal homodyne phase $\theta_{\rm opt}$. It is proportional to the squared optomechanical coupling $g^2$ and therefore, in the situation considered here, it can be treated as a back-action-like term. Here we also have to take into account that stability conditions imply that the squeezing parameter $N$ cannot be too small, that is, one has to impose that $N>N_{\rm min}$, and, in fact, for $N<N_{\rm min}$, the total force noise spectrum of Eq.~\eqref{eq31} becomes negative. Therefore, force noise reduction is maximum close to the stability threshold, close $N = N_{\rm min}$,
which generally depends upon $G/g$, $\Gamma/\gamma_m$, and also $\omega$.
For example, considering $(G-g)/g=10^{-3}$, $(\Gamma-\gamma_m)/\gamma_m=10^{-2}$, and $\omega=\omega_m+4\gamma_m$, the numerical solution of the nonlinear algebraic equation
$S_{F,\theta_{\rm opt}}(\omega)=0$ for $N_{\rm min}$ leads to  $N_{\mathrm{min}}=0.168\,403$.
According to Eq.~\eqref{eq31}, in the case of imperfect CQNC and the resonant cavity driving, the noise reduction brought by the homodyne phase optimization is
\begin{align}
\delta S_{\rm ICQNC}=& -\frac{4g^2}{\kappa\gamma_m}\Bigg[\mathrm{Re}\Bigg(\frac{1+(G/g)^2R(\omega)}{\chi_m(\omega)}\Bigg)\Bigg]^2\notag\\
&\times\frac{|\chi_m(\omega)|^2}{N+1/2+\sqrt{N(N+1)}}, \label{eq32}
\end{align}
which can be rewritten in decibel scale.
\begin{figure}
	\begin{center}
		\includegraphics[width=8cm]{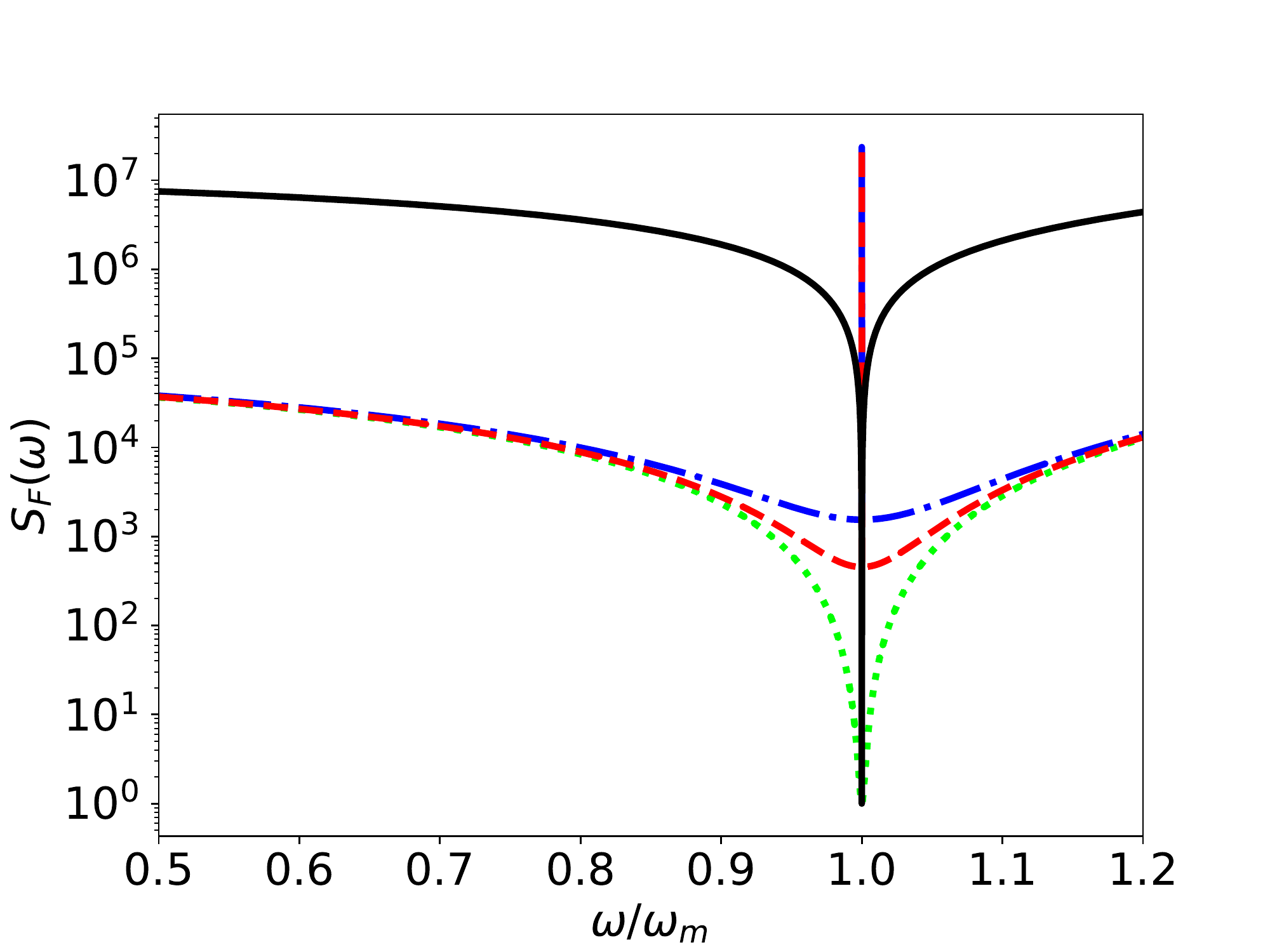}
	\end{center}
	\vspace*{-5mm}
	\caption{(Color online) Spectral density of the force noise versus $\omega/\omega_m$, in the case of the resonant cavity driving $\Delta_c=0$, with an optimized squeezed injected light with $\phi=\phi_{\rm opt}(0)=0$, $|M|=\sqrt{N(N+1)}$, and $N=0.2$. We choose the mismatches between the MO and the effective NMO parameters as $(G-g)/g=10^{-3}$ and $(\Gamma-\gamma_m)/\gamma_m=-0.2$ (blue dash-dotted line for $\theta=0$ and red dashed line for $\theta=\theta_{\rm opt}$). The green dotted line and black solid line correspond to the perfect CQNC and SQL, respectively. The other parameter values are the same as those in Fig.~\ref{fig2}.  }
	\label{fig6}
\end{figure}

\begin{figure}
	\begin{center}
		\includegraphics[width=8cm]{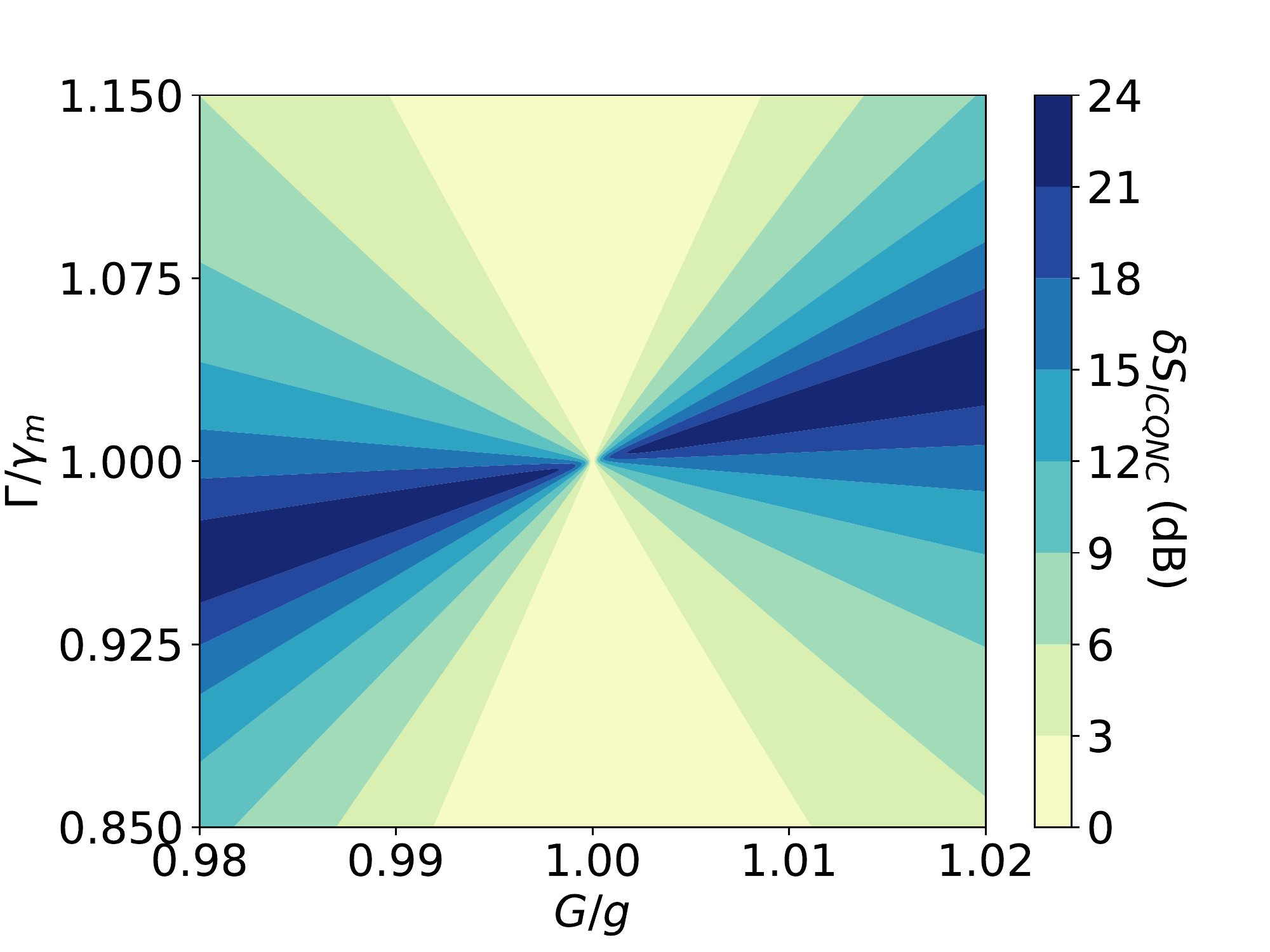}
	\end{center}
	\vspace*{-5mm}
	\caption{(Color online) Noise reduction advantage (in decibels), at off-resonance frequency $\omega=\omega_m-4\gamma_m$ versus the coupling mismatch $G/g$ (horizontal axis) and decay rate mismatch $\Gamma/\gamma_m$ (vertical axis). Here we consider an optimized squeezed injected light with $\phi=\phi_{\rm opt}(0)=0$, $|M|=\sqrt{N(N+1)}$, and $N=0.126$. In addition, we assume that the cavity is driven at resonance $\Delta_c=0$. The other parameters are the same as those in Fig.~\ref{fig2}.}
	\label{fig7}
\end{figure}

\begin{figure}
	\begin{center}
		\includegraphics[width=8cm]{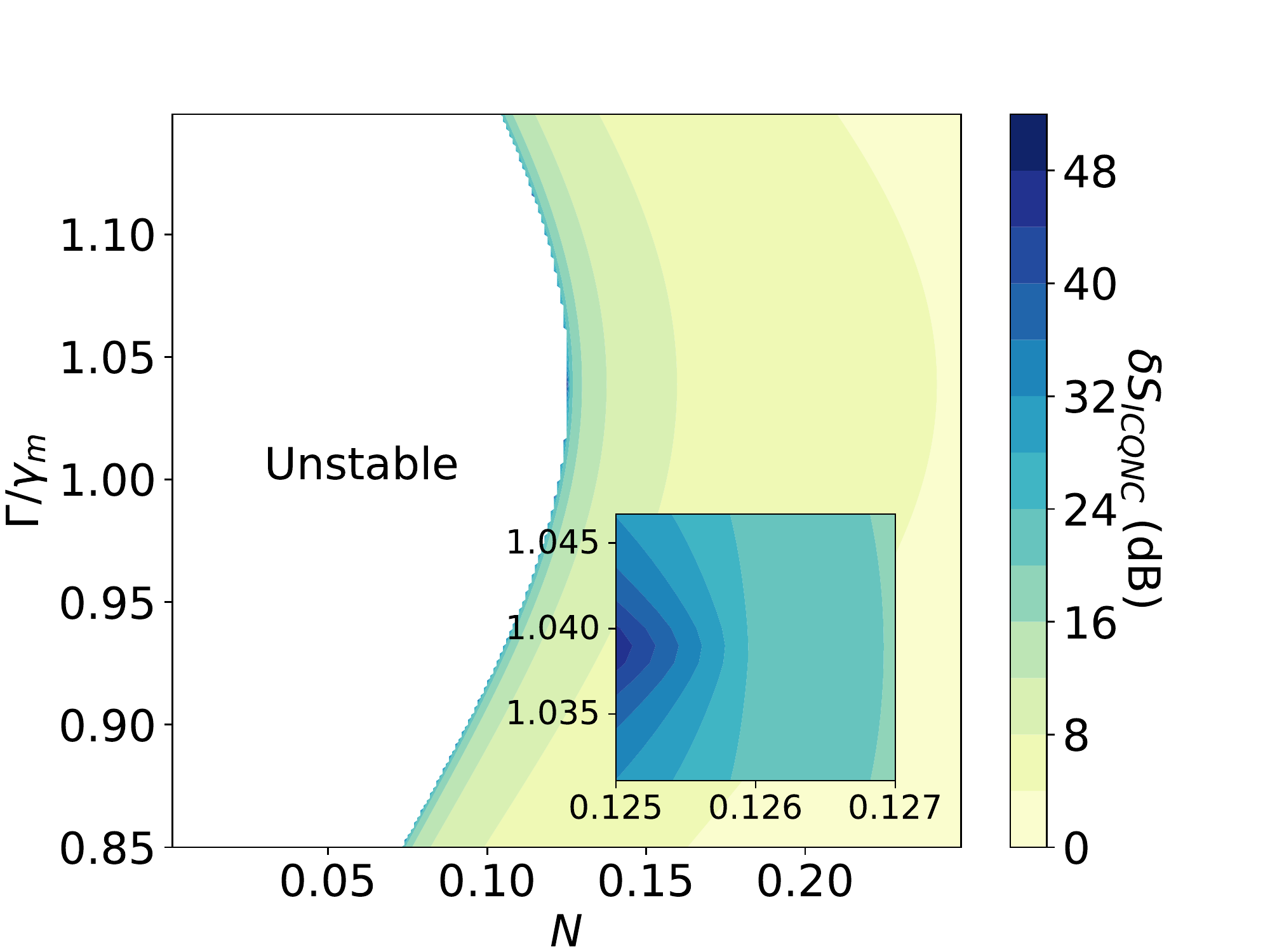}
	\end{center}
	\vspace*{-5mm}
	\caption{ (Color online) Noise reduction advantage (in decibels) at off-resonance frequency $\omega=\omega_m-4\gamma_m$ versus the dissipation rate mismatch $\Gamma/\gamma_m$ (vertical axis) and the squeezing parameter $N$ (horizontal axis) in the case of resonant cavity driving $\Delta_c=0$. The coupling rate mismatch is $(G-g)/g=0.02$ and an optimized squeezed light with $\phi=\phi_{\rm opt}(0)=0$ and $|M|=\sqrt{N(N+1)}$ is considered. The other  parameter values are the same as those in Fig.~\ref{fig2}.}
	\label{fig8}
\end{figure}

\begin{figure}
	\begin{center}
		\includegraphics[width=8cm]{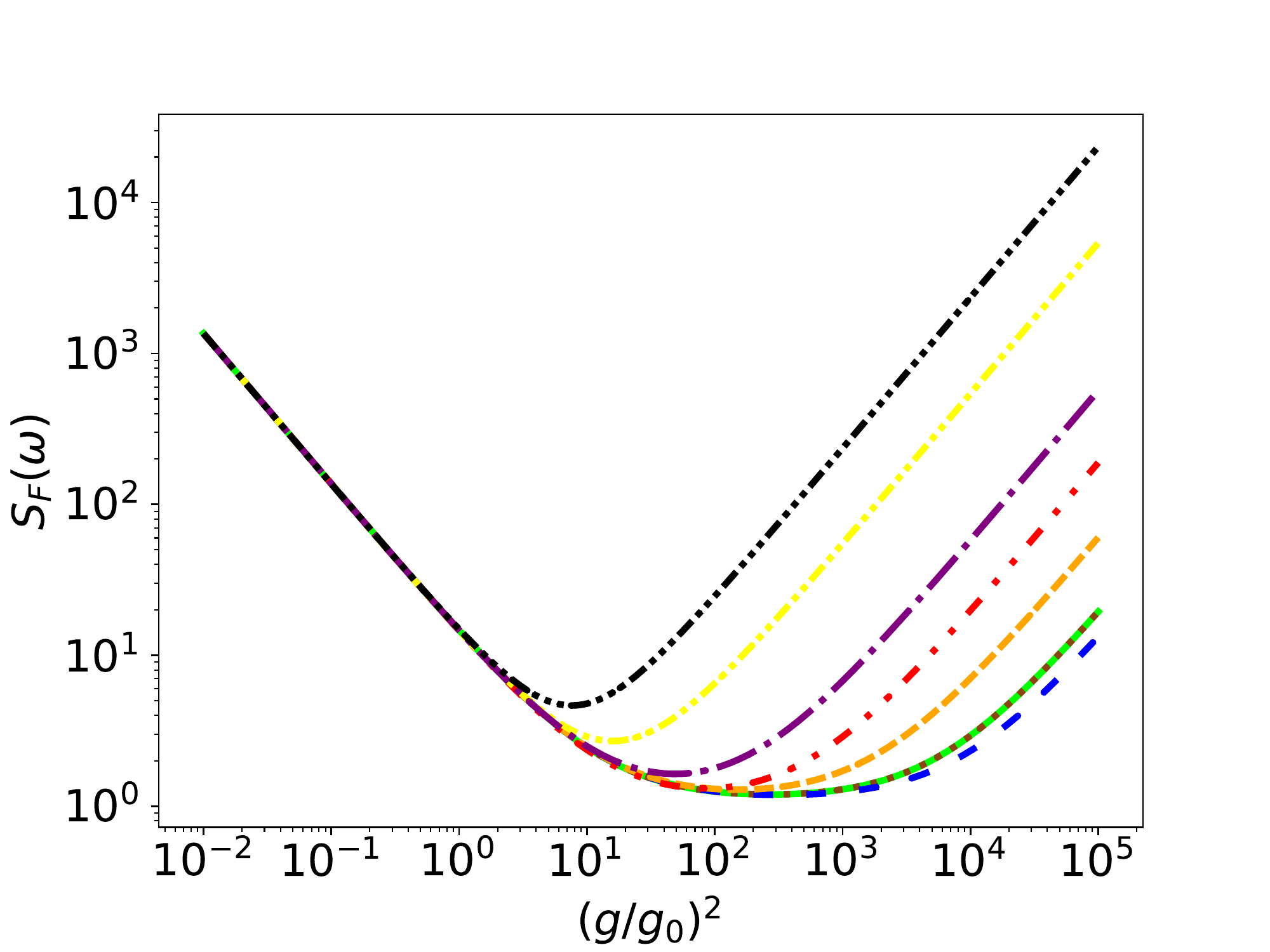}
	\end{center}
	\vspace*{-5mm}
	\caption{(Color online) Force noise spectrum versus $(g/g_0)^2$ at off-resonance frequency $\omega=\omega_m+4\gamma_m$ in the case of resonant cavity driving
$\Delta_c=0$. An optimized squeezed injected light with $\phi=\phi_{\rm opt}(0)=0$, $|M|=\sqrt{N(N+1)}$ and $N=0.1245$ is considered. Different curves correspond to
		$G=g$ and $(\Gamma-\gamma_m)/\gamma_m=0.1$ (green solid  line for $\theta=\theta_{\rm opt}$ and brown dotted line for $\theta=0$),
		$(G-g)/g=0.01$ and $(\Gamma-\gamma_m)/\gamma_m=0.1$ (blue dashed line for $\theta=\theta_{\rm opt}$ and orange densely dashed line for $\theta=0$),
		$(G-g)/g=-(\Gamma-\gamma_m)/\gamma_m=0.1$ (red dash--double-dotted line for $\theta=\theta_{\rm opt}$ and yellow densely dash--double-dotted line for $\theta=0$),
		$(G-g)/g=0.2$ and $(\Gamma-\gamma_m)/\gamma_m=-0.1$ (purple dash-dotted line for $\theta=\theta_{\rm opt}$ and black dash--triple-dotted line for $\theta=0$).
		The other parameter values are the same as those in Fig.~\ref{fig2}.     }
	\label{fig9}
\end{figure}

\begin{figure}
	\begin{center}
		\includegraphics[width=8cm]{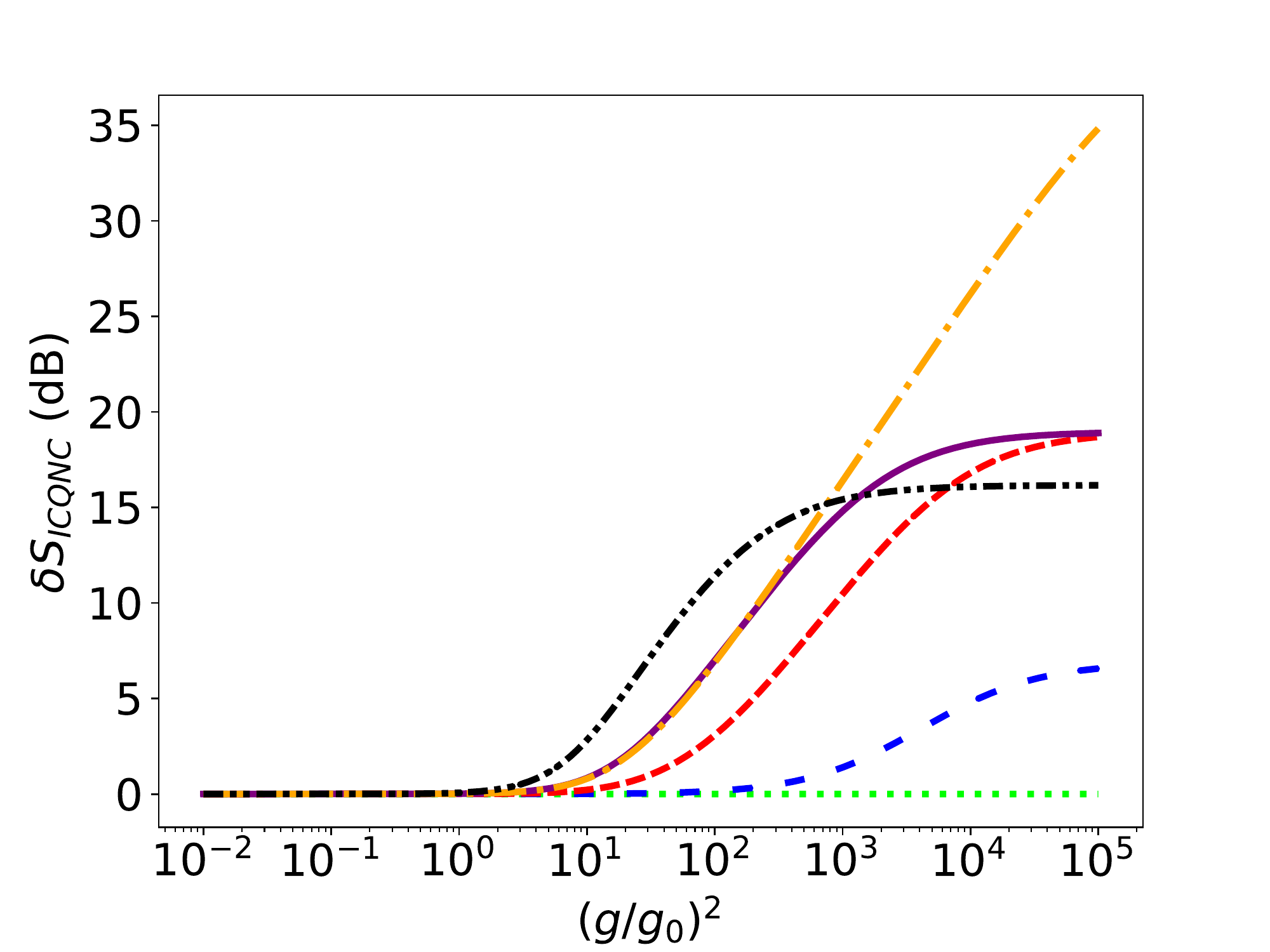}
	\end{center}
	\vspace*{-5mm}
	\caption{(Color online) Noise reduction advantage (in decibels) versus $(g/g_0)^2$ at off-resonance frequency $\omega=\omega_m+4\gamma_m$ in the case of resonant cavity driving
$\Delta_c=0$. An optimized squeezed injected light with $\phi=\phi_{\rm opt}(0)=0$, $|M|=\sqrt{N(N+1)}$, and $N=0.1245$ is considered here. Different curves correspond
to
		$G=g$ and $(\Gamma-\gamma_m)/\gamma_m=0.1$ (green dotted line),
		$(G-g)/g=0.01$ and $(\Gamma-\gamma_m)/\gamma_m=0.1$ ( blue dashed line),
		$(G-g)/g=0.05$ and $\Gamma=\gamma_m$ (red densely dashed line),
		$(G-g)/g=0.1$ and $\Gamma=\gamma_m$ (purple solid line),
		$(G-g)/g=(\Gamma-\gamma_m)/\gamma_m=0.1$ (orange dash-dotted line),
		 and $(G-g)/g=0.2$ and $(\Gamma-\gamma_m)/\gamma_m=-0.1$ (black dash--double-dotted line).
		The other parameter values are the same as those in Fig.~\ref{fig2}.     }
	\label{fig10}
\end{figure}

Figure~\ref{fig6} explicitly shows the force noise reduction associated with the optimization over the phase of the homodyne detected quadrature: We have plotted the spectral density of the force noise versus the frequency and compared the results with the SQL of the force measurement and also
with the case of perfect CQNC. This figure shows that although the noise spectrum under imperfect CQNC exceeds that under perfect CQNC conditions, it still
remains below the SQL in a broad band around the resonance peak. Furthermore, it demonstrates that for $(G-g)/g=10^{-3}$ and $(\Gamma-\gamma_m)/\gamma_m=-0.2$, the variational readout of the cavity output field obviously reduces the force noise spectrum around the resonance frequency $\omega=\omega_m$.
The impact of coupling and decay rate mismatches on the noise reduction advantage is explicitly shown in Fig.~\ref{fig7}. This figure shows that the noise reduction advantage is more sensitive to the coupling rate mismatch than to the decay rate mismatch. For instance, when the coupling rate mismatch is about $0.5\%$ ($G/g\simeq0.995$), a $2.0\%$
variation of the decay rate mismatch (from $\Gamma/\gamma_m=1.02$ to $\Gamma/\gamma_m=1.04$) leads to a nearly $4~\mathrm{dB}$ change in the noise reduction advantage, while a
$2\%$ variation of the coupling rate mismatch, at any decay rate mismatch, is equivalent to at least a $12~\mathrm{dB}$ change in the noise reduction advantage. Furthermore,
this figure demonstrates that when the coupling rates are perfectly matched ($G=g$), the noise reduction advantage is negligible. This is due to the fact that for $G=g$, by
considering the system parameters as in Fig.~\ref{fig6}, we can conclude that (see Appendix \ref{app.B})
\begin{equation}\label{eq33}
\delta S_{\rm ICQNC}\approx-\frac{4g^2}{\kappa\gamma_m}\frac{(1-\Gamma /\gamma_m)^2}{N+1/2+\sqrt{N(N+1)}},
\end{equation}
which is negligible for any values of $\Gamma/\gamma_m$ within the interval considered in Fig.~\ref{fig7}.
Moreover, Fig.~\ref{fig8} shows how the squeezing parameter $N$ affects the noise reduction advantage. This figure shows that $\delta S_{\rm ICQNC}$ increases
as the squeezing parameter $N$ decreases, and eventually it reaches its maximum value. This behavior is expected from Eq.~\eqref{eq32}, which shows that $\delta S_{\rm ICQNC}$ is inversely proportional to $N$.
As it is clear from Fig.~\ref{fig8}, in a large parameter range the noise reduction advantage is less than $16~\mathrm{dB}$, while, as shown in the inset, in a very tiny region close to the instability threshold, with
$0.1250\leq N \leq 0.1256$ and $0.0387\leq(\Gamma-\gamma_m)/\gamma_m\leq 0.0392$, the advantage brought by the optimization over the homodyne phase in reducing the backaction noise can be improved up to $40~\mathrm{dB}$.
As we have already remarked, according to Eq.~\eqref{eq32}, the noise reduction advantage improves as the cavity driving power increases. This behavior is illustrated in Fig.~\ref{fig9}, which compares the force noise
spectrum at the optimal homodyne phase ($\theta=\theta_{\rm opt}$) with that for the standard CQNC ($\theta=0$) studied in Ref.~\cite{motazedifard2016force}.
In the case of perfect coupling rate matching $G=g$, choosing the optimal phase quadrature provides no advantage in reducing the force noise spectrum compared to the standard detection with $\theta = 0$. In contrast, in the absence of perfect matching of the coupling rates, Fig.~\ref{fig9} shows the advantage of variational homodyne readout in reducing the
backaction noise. This result is confirmed in Fig.~\ref{fig10}, in which the noise reduction advantage is plotted versus the cavity driving power. Analogous to
Fig.~\ref{fig9}, this figure shows that the noise reduction advantage increases as the cavity driving power grows. For instance, in the case of
$(G-g)/g=(\Gamma-\gamma_m)/\gamma_m=0.1$, homodyne CQNC leads to the reduction of the force noise up to $35~\mathrm{dB}$ over the standard CQNC in
Ref.~\cite{motazedifard2016force} at sufficiently high driving powers.

\subsubsection{off-resonant case $(\Delta_c\ne 0)$}
\begin{figure}
	\begin{center}
		\includegraphics[width=8cm]{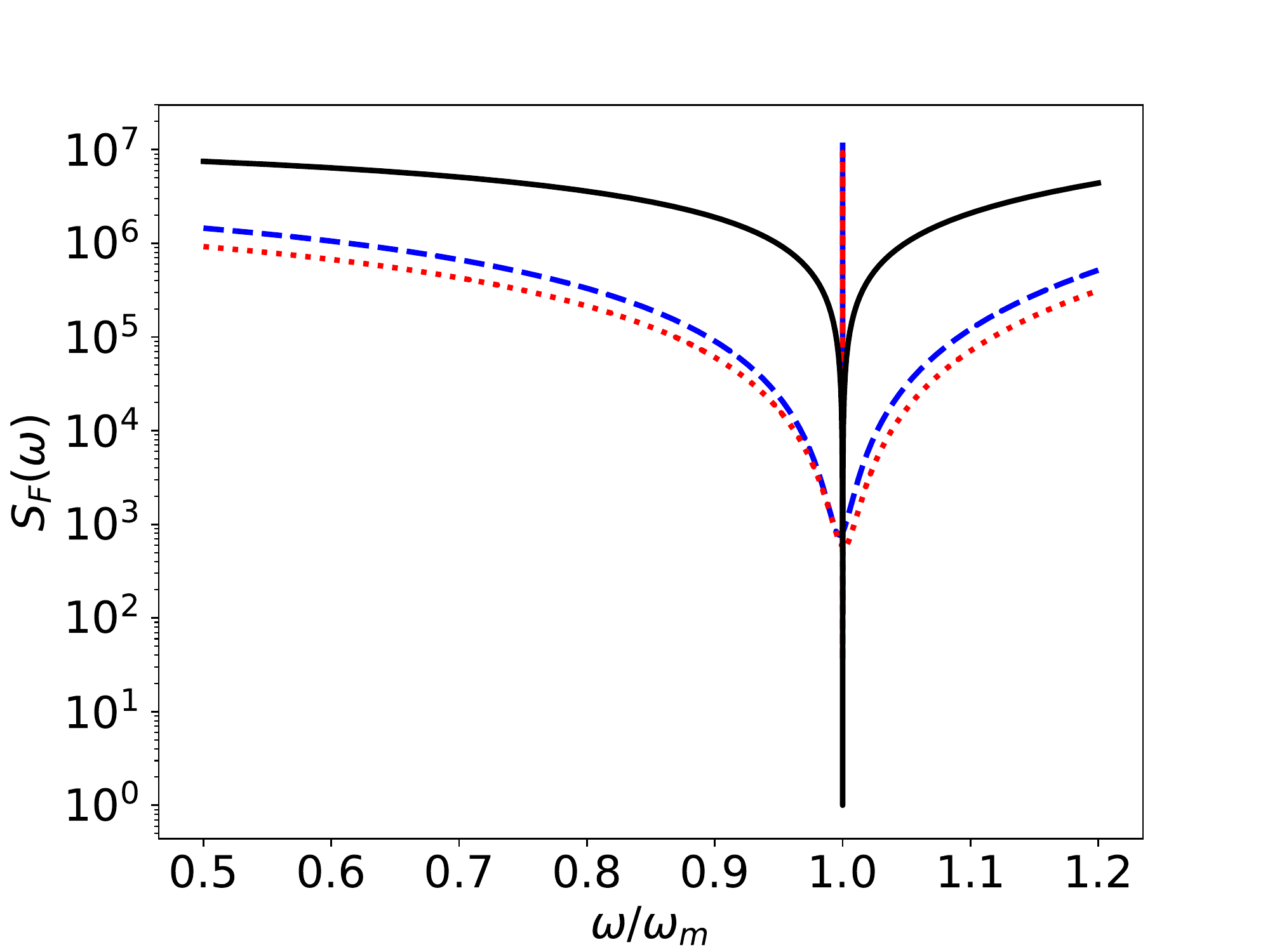}
	\end{center}
	\vspace*{-5mm}
	\caption{(Color online) Spectral density of the force noise $S_F(\omega)$ versus the normalized frequency $\omega/\omega_m$, in the case of the off-resonance cavity
driving with $\Delta_c/\kappa=1$. An optimized squeezed injected light with $|M|=\sqrt{N(N+1)}$ and $N=1$ is considered. Different curves correspond to
	 $(G-g)/g=10^{-5}$ and $\Gamma=\gamma_m$ (blue dashed line for $\theta=0$ and red dotted line for $\theta=\theta_{\rm opt}$);the black solid line corresponds to the SQL. The other parameter values are the same as those in Fig.~\ref{fig2}.      }
	\label{fig11}
\end{figure}

\begin{figure}
	\begin{center}
		\includegraphics[width=8cm]{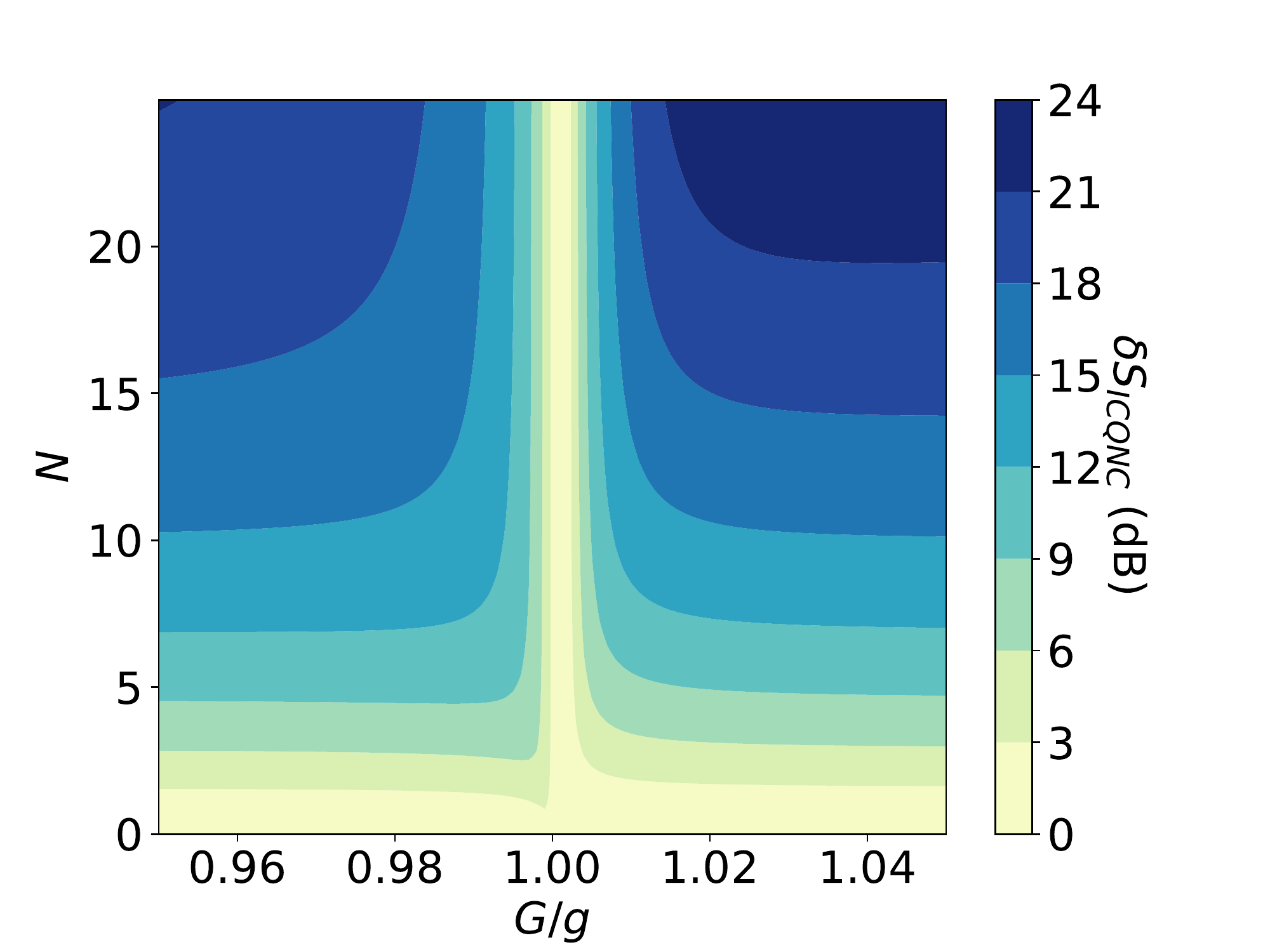}
	\end{center}
	\vspace*{-5mm}
	\caption{(Color online) Noise reduction advantage (in decibels) versus the coupling rate mismatch (horizontal axis) and the parameter $N$ (vertical axis), at off-resonance
frequency $\omega=\omega_m-4\gamma_m$ and in the case of off-resonant cavity driving with $\Delta_c/\kappa=1$. An optimized squeezed injected light with $|M|=\sqrt{N(N+1)}$
is considered here. We also assume that the decay rate mismatch is $(\Gamma-\gamma_m)/\gamma_m=0.2$. The other parameter values are the same as those in Fig.~\ref{fig2}.    }
	\label{fig13}
\end{figure}

\begin{figure}
	\begin{center}
		\includegraphics[width=8cm]{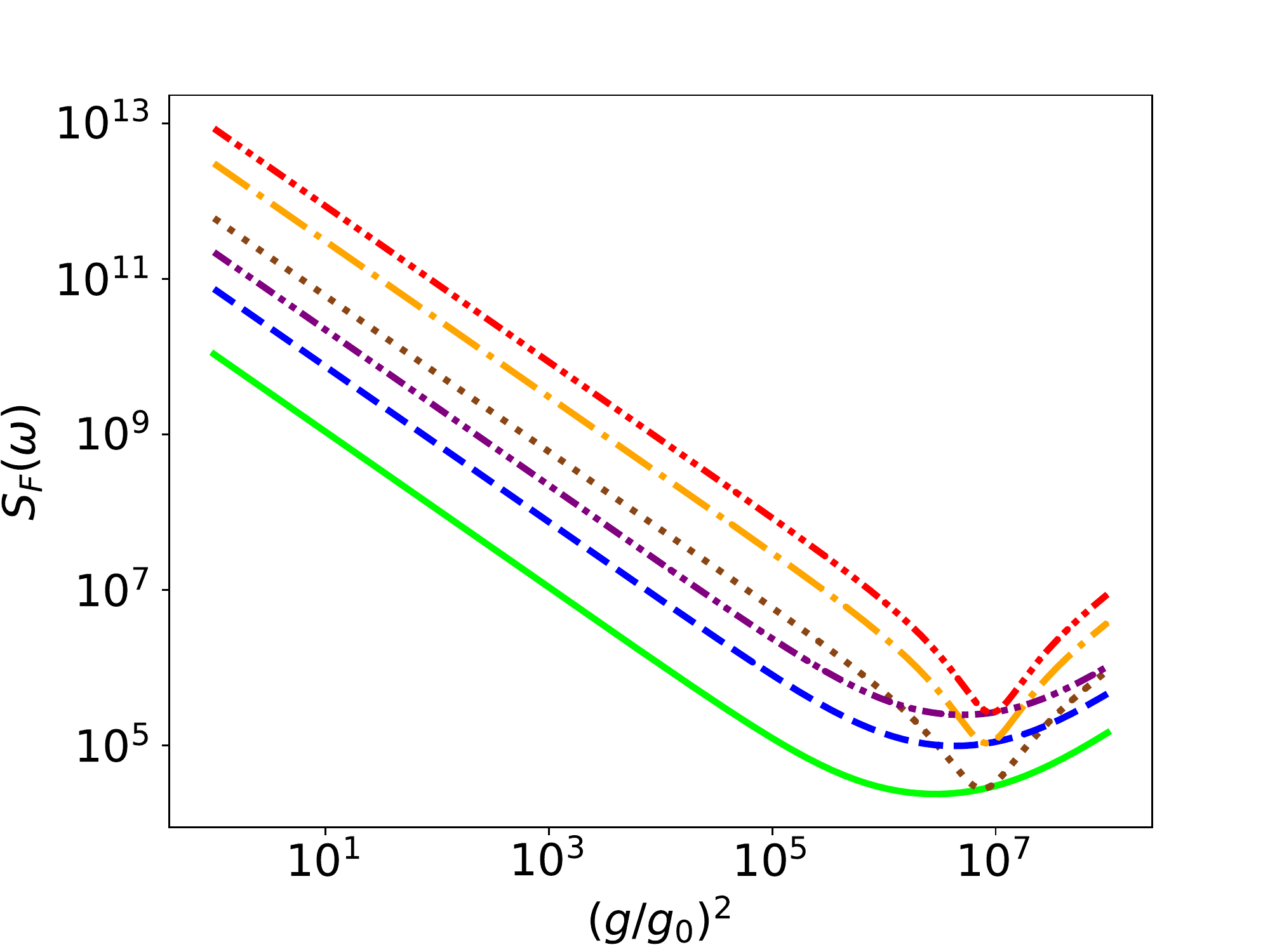}
	\end{center}
	\vspace*{-5mm}
	\caption{(Color online) Force noise spectrum versus $(g/g_0)^2$ at off-resonance frequency $\omega=\omega_m-4\gamma_m$ and in the case of off-resonant cavity driving
$\Delta_c/\kappa=1$. An optimized squeezed vacuum injected light with $\phi=\phi_{\rm opt}(1)$, $|M|=\sqrt{N(N+1)}$, and $N=20$ is considered. Different curves
correspond to
	$(G-g)/g=0.05$ and $(\Gamma-\gamma_m)/\gamma_m=0.1$ (green solid line for $\theta=\theta_{\rm opt}$ and brown dotted line for $\theta=0$),
	$(G-g)/g=0.1$ and $(\Gamma-\gamma_m)/\gamma_m=0.15$ (blue dashed line for $\theta=\theta_{\rm opt}$ and orange dash-dotted line for $\theta=0$), and
	$(G-g)/g=0.15$ and $(\Gamma-\gamma_m)/\gamma_m=0.2$ (purple dash--double-dotted line for $\theta=\theta_{\rm opt}$ and red dash--triple-dotted line for $\theta=0$). The other parameter values are as those in Fig.~\ref{fig2}.  }
	\label{fig14}
\end{figure}

\begin{figure}
	\begin{center}
		\includegraphics[width=8cm]{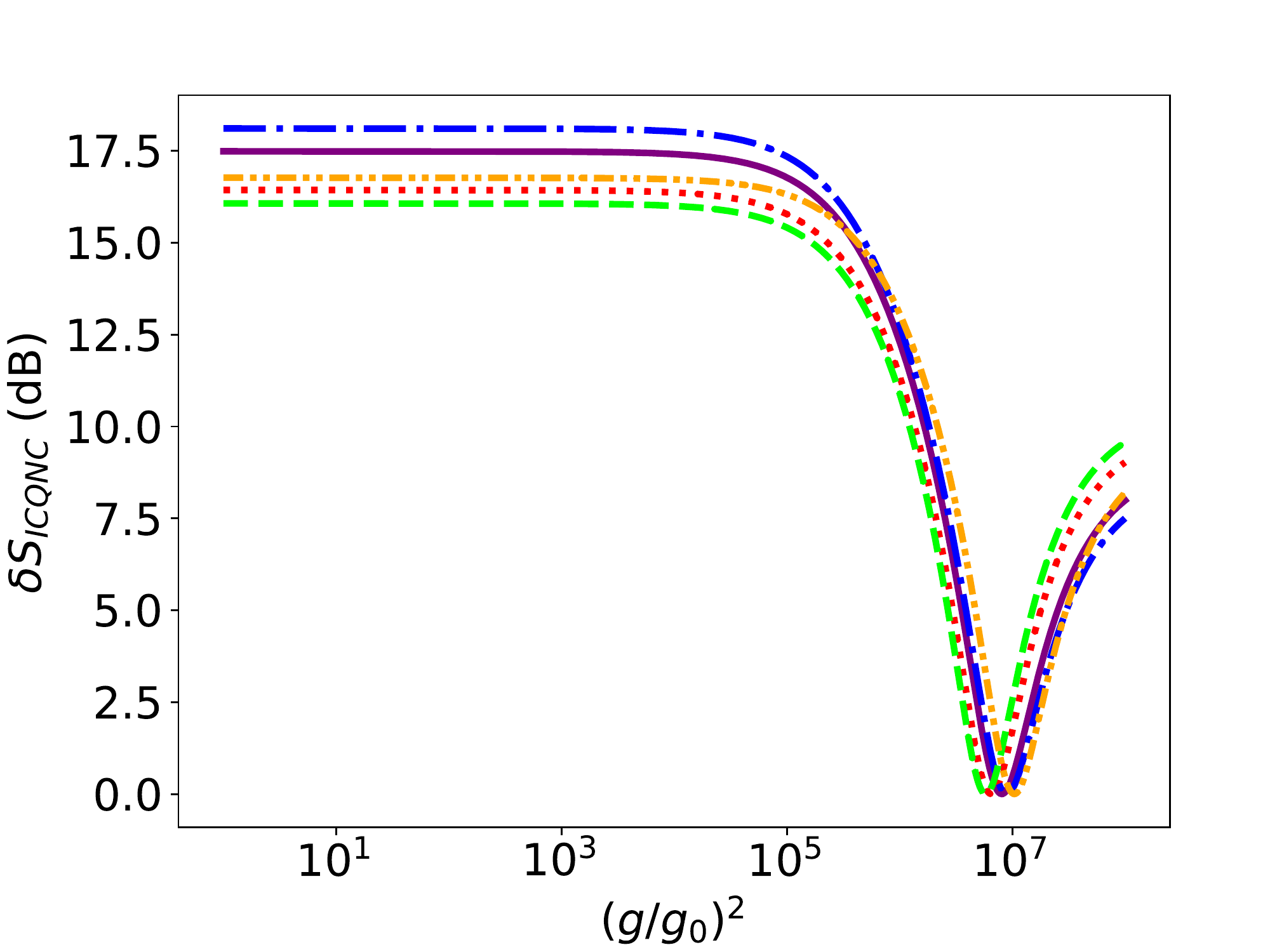}
	\end{center}
	\vspace*{-5mm}
	\caption{(Color online) Noise reduction advantage (in decibels) versus $(g/g_0)^2$ at off-resonance frequency $\omega=\omega_m-4\gamma_m$ and in the case of off-resonant cavity driving with $\Delta_c/\kappa=1$. An optimized squeezed vacuum injected light with $\phi=\phi_{\rm opt}(1)$, $|M|=\sqrt{N(N+1)}$, and $N=20$ is considered. Different curves correspond to
	$(G-g)/g=0.05$ and $(\Gamma-\gamma_m)/\gamma_m=0.1$ (purple solid line),
	$(G-g)/g=-0.05$ and $(\Gamma-\gamma_m)/\gamma_m=-0.1$ (red dotted line),
	$(G-g)/g=0.1$ and $(\Gamma-\gamma_m)/\gamma_m=0.2$ (blue dash-dotted line),
	$(G-g)/g=-0.1$ and $(\Gamma-\gamma_m)/\gamma_m=-0.2$ (green dashed line), and
	$(G-g)/g=0.2$ and $(\Gamma-\gamma_m)/\gamma_m=0.3$ (orange dash--double-dotted line). The other parameter values are the same as those in Fig.~\ref{fig2}.}
	\label{fig15}
\end{figure}

In this section we consider imperfect CQNC conditions when the cavity is driven off-resonance $(\Delta_c\ne0)$. We derived the most general form of the force noise spectrum earlier in Sec.~\ref{subsec4A}. Here we minimize the force noise spectrum $S_{F,\theta}(\omega)$ given by Eq.~\eqref{eq12} over the
parameter $\theta$ and obtain
\begin{equation}\label{eq35}
	S_{F,\theta_{\rm opt}}^{\rm min}(\omega)=S_{F,\theta=0}(\omega)-\frac{{K^\prime}^2}{4L^\prime},
\end{equation}
for the minimized force noise spectrum, in which
\begin{equation}\label{eq36}
\tan\theta_{\rm opt}=-\frac{K^\prime/2L^\prime}{1-y(K^\prime/L^\prime)}
\end{equation}
is the optimum phase of the homodyne detection, and
\begin{subequations}
	\begin{align}
	L^\prime=&\frac{2}{g^2\kappa\gamma_m|\chi_a^\prime(\omega)|^2|\chi_m(\omega)|^2}\bigg\{\mu_{+}\bigg(N+\frac{1}{2}\bigg) + \mu_{-}\mathrm{Re}M \notag\\
	&\,+ 2y\mathrm{Im}(M) \bigg\}, \label{eq37a}\\
	K^\prime=&\frac{1}{g^2\gamma_m|\chi_m(\omega)|^2}\Bigg\{4\Delta_c\mathrm{Re}\Bigg[\frac{Z(\omega)}{\chi_a^{\prime}(-\omega)} \Bigg]\mathrm{Re}M \notag\\
	&\, +\mathrm{Im}\bigg[(2i\mathrm{Im}M+1)\bigg(\frac{4y^2}{\chi_a^\prime(\omega)} -\frac{1}{\kappa|\chi_a^\prime(\omega)|^2}\notag\\
	&\, + \frac{1-\kappa Z(-\omega)}{\chi_a^\prime(-\omega)} \bigg) \bigg] \Bigg\}\notag\\
	&\, - \frac{2}{\kappa\gamma_m}\Bigg\{2\mathrm{Re}\bigg[\frac{1+(G/g)^2R(\omega)}{\chi_m(\omega)\chi_a^\prime(\omega)}\bigg](N+\frac{1}{2}+\mathrm{Re}M)\notag\\
	&\, +2y\mathrm{Im}\bigg[(2i\mathrm{Im}M+1)\frac{1+(G/g)^2R(\omega)}{\chi_m(\omega)\chi_a^\prime(\omega)}\bigg] \Bigg\}.\label{eq37b}
	\end{align}
	\label{eq37-all-lines}
\end{subequations}
Similar to the previous cases, the advantage of our optimized CQNC scheme in reducing the force noise over the standard CQNC is determined by the noise reduction advantage
\begin{equation}\label{eq38}
	\delta S_{\rm ICQNC}=-\frac{{K^\prime}^2}{4L^\prime},
\end{equation}
which implies that our scheme is advantageous as long as $L^\prime$ is positive.
In Fig.~\ref{fig11} we compare the force noise spectrum in our scheme with that considered in \cite{motazedifard2016force} and with the SQL of the force measurement.
According to this figure, we see that the variational homodyne CQNC with optimized quadrature phase mitigates the effect of imperfect matching conditions.
Our numerical calculations demonstrate that the noise reduction advantage is more sensitive to the coupling rate mismatch rather than to the decay rate mismatch such that for $y=1$ and the optimum parameters for squeezed injected light such as $\phi=\phi_{\rm opt}(1)$ and $|M|=\sqrt{N(N+1)}$, with $N=10$, the noise reduction advantage is suddenly increased by $16~\mathrm{dB}$ as the coupling rate mismatch grows.
Figure~\ref{fig13} shows that noise reduction is amplified as $N$ increases, such that for $(\Gamma-\gamma_m)/\gamma_m=0.2$ and $(G-g)/g>0.015$ the noise reduction
advantage can be increased by $15~\mathrm{dB}$ for $10<N<25$.
We also investigate how the cavity driving power affects the noise reduction advantage in Figs.~\ref{fig14} and \ref{fig15}, which demonstrate the
variation of the force noise spectrum and the noise reduction advantage versus the cavity driving power, respectively.
According to Fig.~\ref{fig15}, although the noise reduction remains nearly constant in a wide range of driving powers, i.e., $(g/g_0)^2<10^5$, it decreases as the driving
power grows and eventually vanishes at a specific driving power which is determined by solving the equation $K^\prime=0$ [Eq.~\eqref{eq37b}] for $(g/g_0)^2$. For example,
when the cavity detuning is $\Delta_c/\kappa=1$, and the coupling rate and the decay rate mismatches are $(G-g)/g=0.1$ and $(\Gamma-\gamma_m)/\gamma_m=0.2$ , respectively,
the homodyne--CQNC with an optimized phase angle $\theta_{\rm opt}$ shows a $18~\mathrm{dB}$ advantage in reduction of the force noise over the standard CQNC.


\section{\label{sec5} Sensitivity, SNR, and signal response amplification}
In this section it is worthwhile to clarify the advantage of the variational homodyne readout CQNC over the standard CQNC in sensitivity, SNR, and also in signal-response amplification. We follow the approach introduced in Refs.~\cite{motazedifard2021ultraprecision,mehryMagneticsensing2020} to obtain these quantities.
Based on Eq.~\eqref{eq9}, there are two contributions to the estimated external force obtained from the experimental signal: The first is the external classical force exerted on the MO, i.e., $F_{\rm ext}$, and the second one is the added force noise $\hat{F}_{N,\theta}$, given by Eq.~\eqref{eq10}, which yields the spectral density of the added force noise $S_{F,\theta}(\omega)$ obtained in Eq.~\eqref{eq12}. The SNR $\Re$ is defined as the ratio of the signal to the variance of the noise, i.e., the square root of the
$S_{F,\theta}(\omega)$ \cite{bemani2021force}, as
\begin{equation}\label{eq39}
	\Re \equiv \frac{|F_{\rm ext}(\omega)|}{\sqrt{S_{F,\theta}(\omega)}}.
\end{equation}
The sensitivity of a force sensor $\mathcal{S}(\omega)$ is defined as the minimum magnitude of the force signal $F_{\rm ext}(\omega)$ for which the $\rm SNR$ of the force
measurement becomes one, i.e., $\Re=1$. From Eq.~\eqref{eq39}, the sensitivity of the force sensor considered here is given by
\begin{equation}\label{eq40}
	\mathcal{S}(\omega)=\sqrt{S_{F,\theta}(\omega)}.
\end{equation}
Let us now examine the advantage of the present scheme in which we optimize the phase of the detected quadrature compared to the standard CQNC studied in
Ref.~\cite{motazedifard2016force}, in improving the force measurement sensitivity. We compare the results at a given frequency, in this case at $\omega=\omega_m-0.2 Q_m \gamma_m=0.8 \omega_m$. As mentioned in
Sec.~\ref{sec4}, under perfect CQNC conditions and at resonant cavity driving, the noise reduction advantage vanishes and therefore there is no sensitivity improvement. However, for nonzero cavity detuning, e.g., $\Delta_c/\kappa=\frac{1}{2}$ and the squeezing parameter $N=10$ (and still perfect CQNC), the force sensitivity with the optimized phase angle $\theta=\theta_{\rm opt}$ is $1.759\times10^{-19}~\mathrm{N}\,\mathrm{Hz}^{-1/2}$, while for $\theta=0$, i.e., standard CQNC, it is $6.194\times10^{-19}~\mathrm{N}\,\mathrm{Hz}^{-1/2}$, i.e., an almost $71.60 \%$ improvement.
In the case of imperfect CQNC conditions, such that $(G-g)/g=0.01$ and $(\Gamma-\gamma_m)/\gamma_m=0.1$, when $\Delta_c=0$ and the squeezing parameter is $N=0.125$, the
force sensitivity for $\theta=\theta_{\rm opt}$ is $6.126\times10^{-17}~\mathrm{N}\,\mathrm{Hz}^{-1/2}$, while for $\theta=0$ is $6.514\times10^{-17}~\mathrm{N}\,\mathrm{Hz}^{-1/2}$, which is about a $5.95 \%$ improvement in sensitivity.
It should be noted that the squeezing parameter $N=0.125$ considered in this case corresponds just to the stability threshold value $N_{\rm min}$ for which the noise reduction advantage is maximum (see Fig.~\ref{fig8}).
Moreover, for the case in which the imperfect CQNC holds such that $(G-g)/g=0.01$ and $(\Gamma-\gamma_m)/\gamma_m=0.1$, when the cavity detuning is $\Delta_c=\kappa$ and the squeezing parameter is $N=25$, the force sensing sensitivity is $3.494\times10^{-16}~\mathrm{N}\,\mathrm{Hz}^{-1/2}$ by considering  $\theta=\theta_{\rm opt}$, while for $\theta=0$ it is $3.804\times10^{-16}~\mathrm{N}\,\mathrm{Hz}^{-1/2}$, which is equivalent to an $8.13\%$ enhancement.

We now investigate the effect of the variational homodyne detection scheme on the signal-response amplification. From Eq.~\eqref{eq9}, one can rewrite the cavity homodyne-output spectrum as
\begin{equation}
S_{P_{a,\theta}}^{\rm out}(\omega) = R_c(\omega,\theta)S_{F,\theta}(\omega),
\end{equation}
in which
\begin{equation}\label{eq42}
R_c(\omega,\theta) \equiv g^2\kappa\gamma_m|\chi_a^\prime(\omega)|^2|\chi_m(\omega)|^2[u_{\theta}]^2,
\end{equation}
is the signal-response power. The homodyne-output signal can be amplified only if $R_c(\omega,\theta)>1$ (signal amplification condition)
\cite{motazedifard2021ultraprecision}.
We are interested in examining the output signal amplification in regimes where the force noise reduction occurs simultaneously. In particular, we want to investigate whether the homodyne detection with an optimized phase angle is able to simultaneously reduce the force noise and amplify the output signal, compared to the standard case with $\theta = 0$.
Note that the signal-response power for this latter case is obtained by setting $\theta=0$ in Eq.~\eqref{eq42}. Since $u_{\theta=0}=1$, we conclude that
in the standard case of phase detection we have $R_c^{\rm st}(\omega)=g^2\kappa\gamma_m|\chi_a^\prime(\omega)|^2|\chi_m(\omega)|^2$. It is
worth defining a parameter to describe the advantage of the variational detection scheme over the standard one in amplification of the output signal as
\begin{align}
\mathcal{R}&\equiv \frac{R_c(\omega,\theta)}{R_c^{\rm st.}(\omega)}=[u_{\theta}]^2,\label{eq43}
\end{align}	
which we call the signal improvement. It is obvious that $\mathcal{R}>1$ (signal improvement condition) denotes the regime where the variational homodyne detection scheme provides an improved signal, $\mathcal{R}=1$ corresponds to the regime in which variational homodyne detection and phase detection schemes are equivalent in terms of the output signal power, and in the regime where $\mathcal{R}<1$ the variational homodyne detection scheme attenuates the signal.
As described in Sec.~\ref{sec4}, the noise reduction advantage occurs at $ \theta=\theta_{\rm opt}$. Therefore, the variational homodyne CQNC scheme simultaneously shows a noise reduction advantage and provides an improved signal if and only 
if $\mathcal{R}\vert_{\theta_{\rm opt}} = [u_{\theta_{\rm opt}}]^2 > 1 $.

\begin{figure}
	\begin{center}
		\includegraphics[width=8cm]{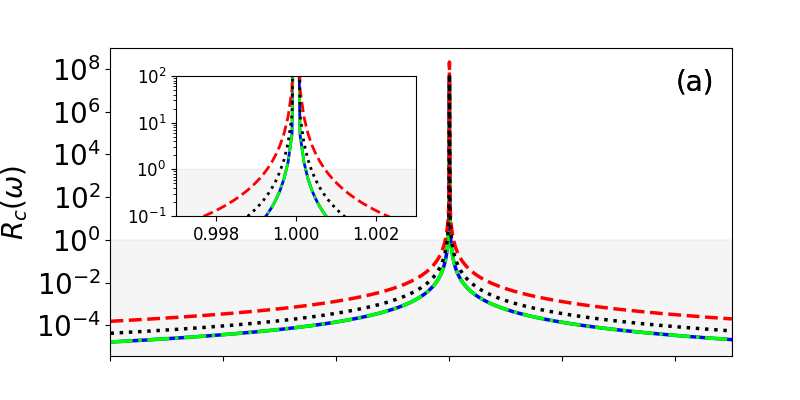}
	\end{center}
	\vspace*{-11mm}
	\begin{center}
		\includegraphics[width=8cm]{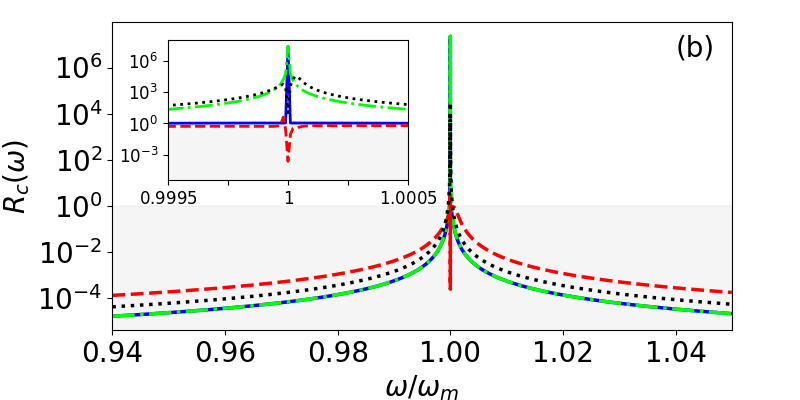}
	\end{center}
	\vspace*{-5mm}
	\caption{(Color online) Power of signal-response versus $\omega/\omega_m$ under the (a) perfect, and (b) imperfect CQNC conditions.
		Different curves in (a) correspond to $\Delta_c=\kappa$ (red dashed line for $\theta=\theta_{\rm opt}$ and black dotted line for $\theta=0$) and $\Delta_c=0$ (blue solid line for $\theta=\theta_{\rm opt}$ and green dash-dotted line for $\theta=0$).
		In (b) we consider $(G-g)/g=10^{-4}$ and $(\Gamma-\gamma_m)/\gamma_m=0.01$ for coupling rate and decay rate mismatches. Different curves in (b) refer to $\Delta_c=0$ (blue solid line for $\theta=\theta_{\rm opt}$ and green dash-dotted line for $\theta=0$) and $\Delta_c=\kappa$ (red dashed line for $\theta=\theta_{\rm opt}$ and black dotted line for $\theta=0$). In both cases, we consider an optimized vacuum injected light with $\phi=\phi_{\rm opt}(y)$, $|M|=\sqrt{N(N+1)}$, and $N=10$. The other parameter values are the same as those in Fig.~\ref{fig2}.}
	\label{fig16}
\end{figure}
\begin{figure}
	\begin{center}
		\includegraphics[width=8cm]{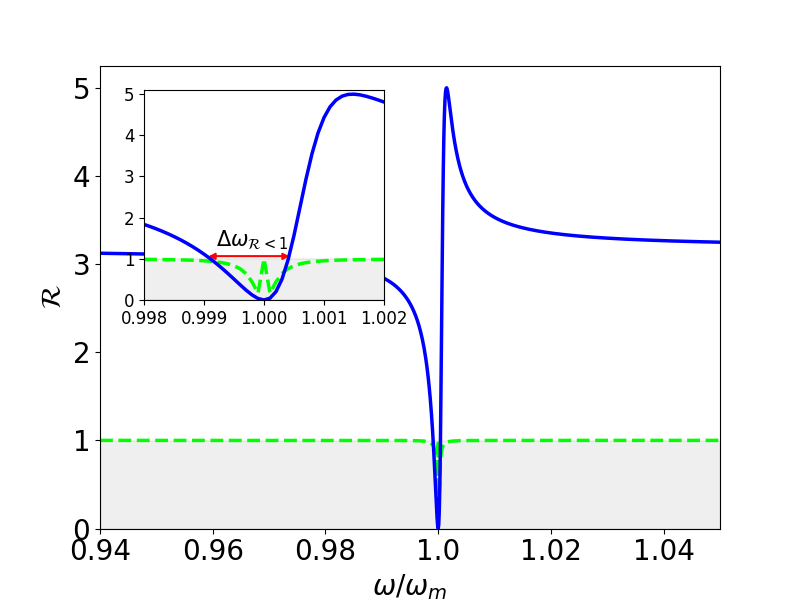}
	\end{center}
	\vspace*{-5mm}
	\caption{(Color online) Signal amplification advantage versus $\omega/\omega_m$,  considering the imperfect CQNC conditions with $(G-g)/g=10^{-4}$ and
		$(\Gamma-\gamma_m)/\gamma_m=0.1$. Different curves correspond to $\Delta_c=0$ (green dashed line) and $\Delta_c=\kappa$ (blue solid line). The other parameter values are the same as those in Fig.~\ref{fig16}.}
	\label{fig17}
\end{figure}

Figure~\ref{fig16} shows the signal-response power $R_c(\omega,\theta)$ versus frequency $\omega/\omega_m$.
We recall that under perfect CQNC conditions and zero detuning ($\Delta_c=0$), the optimized quadrature for which the force noise spectrum is minimized is the phase quadrature, i.e., $\theta_{\rm opt}=0$. Consequently, the variational homodyne detection scheme shows no advantage over the standard CQNC at all frequencies in terms of the signal-response power, i.e., $R_c(\omega,\theta_{\rm opt})=R_c^{\rm st}(\omega)$. We can conclude that under the perfect CQNC conditions and resonant cavity driving, the variational homodyne CQNC scheme does not provide an improved signal, i.e., $\mathcal{R}=1$. In this case, both variational homodyne detection and phase detection schemes show signal amplification in a narrow band around the resonance frequency $\omega=\omega_m$ [see Fig.~\ref{fig16}(a), blue solid and green dash-dotted lines]. In this case, the amplification bandwidth is $B_{\rm CQNC}^{\rm on}=4.86\times10^3\gamma_m$.

When the perfect CQNC conditions hold and the cavity is driven off-resonantly with $\Delta_c=\kappa$, the variational homodyne CQNC strategy provides an improved signal at all frequencies so that  $\mathcal{R}\simeq 3.638$. Although signal improvement, i.e., $\mathcal{R}>1$, occurs at all frequencies, signal amplification, i.e., $R_c(\omega, \theta)>1$, is only observed in a narrow band around the mechanical resonance frequency $\omega \simeq \omega_m$ [see Fig.~\ref{fig16}(a), red dashed and black dotted lines]. The amplification bandwidth $B_{\rm CQNC}$ for the variational homodyne CQNC scheme with the optimized phase $\theta_{\rm opt}$ is $B_{\rm CQNC}^{\rm hom, off}=1.45\times10^4\gamma_m$, while for the standard phase detection scheme with $\theta=0$ it is $B_{\rm CQNC}^{\rm st, off}=7.64\times10^3\gamma_m$. Interestingly, over a region with the bandwidth of $B_{\rm CQNC}$ around the mechanical resonance frequency $\omega_m$, our optimization strategy shows simultaneous noise reduction advantage and signal amplification.

Under imperfect CQNC conditions, the situation is a little more complicated. We consider $(G-g)/g=10^{-4}$ and $(\Gamma-\gamma_m)/\gamma_m=0.01$ for the coupling
rate and decay rate mismatch, respectively. First, in the case of on-resonance cavity driving ($\Delta_c=0$), Fig.~\ref{fig16}(b) illustrates that the variational homodyne detection scheme shows signal amplification, i.e., $R_c(\omega, \theta_{\rm opt.})>1$, in an extremely narrow bandwidth of $B_{\rm ICQNC}^{\rm hom, on}=1.51\times10^3\gamma_m$ around the resonance frequency $\omega_m$ (blue solid line), while in this case the amplification bandwidth for the standard phase detection scheme is $B_{\rm ICQNC}^{\rm st, on}=4.74\times10^{3}\gamma_m$ (green dash-dotted line). Figure~\ref{fig17} shows the signal improvement for the case of imperfect CQNC conditions with $(G-g)/g=10^{-5}$ and $(\Gamma-\gamma_m)/\gamma_m=0.1$. According to this figure, when the cavity is driven on-resonance (dashed green line), the variational homodyne detection scheme neither improves nor weakens the cavity signal at all frequencies ($\mathcal{R}=1$), except in a bandwidth of $1.78\times10^4\gamma_m$ around the resonance frequency $\omega_m$. Within this narrow frequency band, the variational homodyne detection scheme provides a weaker signal compared to the phase detection scheme, i.e., $\mathcal{R}<1$. As a result, in the case of imperfect CQNC conditions and on-resonance cavity driving, the variational homodyne detection scheme proposed here shows no advantage over the standard phase detection scheme considered in \cite{motazedifard2016force} in terms of signal amplification and improvement.

In the case of imperfect CQNC conditions and off-resonance cavity driving with $\Delta_c=\kappa$, Fig.~\ref{fig16}(b) demonstrates that in the variational homodyne detection with $\theta=\theta_{\rm opt}$, signal amplification, i.e., $R_c(\omega, \theta_{\rm opt})>1$, does not occur at all, except in an extremely narrow band around $\omega=\omega_m-21.8\gamma_m$ with a bandwidth of $B_{\rm ICQNC}^{\rm hom, off}=13.8\gamma_m$. One should zoom in to be able to observe this miniature amplification.
In contrast, the standard  phase detection scheme leads to signal amplification, i.e., $R_c^{\rm st}(\omega)>1$, within a bandwidth of $B_{\rm ICQNC}^{\rm st, off}=7.61\times10^3\gamma_m$ around the mechanical resonance frequency $\omega_m$.
On the other hand, Fig.~\ref{fig17} illustrates that our variational homodyne detection scheme provides an improved signal, i.e., $\mathcal{R}>1$, over the standard phase detection scheme at all frequencies, except in an asymmetrical frequency band with the bandwidth of $\Delta\omega_{R<1}=1.30\times10^4\gamma_m$ around the mechanical resonance frequency $\omega_m$ (blue solid line). Furthermore, according to this figure, we observe a peak in signal improvement at $\omega=1.0014\omega_m$ with $\mathcal{R}=5$, while in other frequencies (except in $\Delta\omega_{R<1}$) the signal improvement is approximately $3.18$. We conclude that in the case of imperfect CQNC conditions and off-resonant cavity driving, the proposed strategy of variational homodyne detection scheme provides an improved signal at all frequencies (except in $\Delta\omega_{R<1}$) versus the standard phase detection scheme considered in \cite{motazedifard2016force}. However, we showed that the variational homodyne detection strategy shows no signal amplification, while the standard phase detection scheme results in signal amplification over a region with the bandwidth of $B_{\rm ICQNC}^{\rm st, off}$ (which is smaller than $\Delta\omega_{R<1}$) around $\omega_m$.

\section{\label{sec6} summary, discussion and outlook}
In this paper we have studied the effect of the optimization of the local phase oscillator in the homodyne detection on the reduction of the backaction noise in force sensing based on CQNC in a hybrid OMS consisting of an optomechanical cavity equipped with an atomic ensemble as a NMO.
We have shown that when the parameters of the atomic ensemble are chosen appropriately, it behaves effectively as a NMO, whose interaction with the intracavity field creates an antinoise path to the system dynamics, which allows the cancellation of the backaction noise of the MO \cite{motazedifard2016force,motazedifard2021ultraprecision}. We have also investigated the advantage of the proposed detection strategy in enhancing force sensing sensitivity and amplification of the signal-response power.

Under CQNC conditions, that is, the perfect matching of atomic and mechanical parameters, the backaction noise of the atomic ensemble cancels
perfectly the mechanical one. Moreover, the cavity field shot noise can be suppressed by injecting a squeezed vacuum light into the cavity field. Under these circumstances,
force detection is limited only by atomic noise, which is the price to pay for the realization of this scheme.
In this work, we demonstrated that a variational homodyne measurement of the cavity output field where the phase of the local oscillator is optimized allows further reduction of the force noise, and improves the force detector sensitivity. We focused on two different situations, cases when perfect CQNC conditions hold and when they do not.

When perfect CQNC conditions hold and the cavity field is driven on-resonance ($\Delta_c=0$), no advantage is obtained in terms of noise reduction and signal improvement compared to the standard case where the phase quadrature is measured. On the other hand, in the off-resonance case ($\Delta_c\ne0$), one has a remarkable reduction of the force noise
in a broad band around the mechanical resonance frequency $\omega_m$ up to $40~\mathrm{dB}$ (see Fig.~\ref{fig3}) and at the same time, an improvement of the signal-response power by a factor of $\mathcal{R}=3.638$ occurs.
In addition, the noise reduction increases with increasing squeezing parameter $N$ and with decreasing driving field strength. Furthermore, in this case (perfect CQNC conditions and off-resonant cavity driving), the variational homodyne CQNC scheme broadens the frequency bandwidth over which the signal amplification occurs, i.e., $R_c(\omega, \theta_{\rm opt})>0$.

Then we demonstrated that under the imperfect CQNC conditions, when the cavity is resonantly driven ($\Delta_c=0$), the optimized homodyne measurement of the proper cavity output field quadrature reduces the force noise in a wide frequency range around the mechanical resonance frequency $\omega_m$.
In this case, the noise reduction is more sensitive to the coupling rate mismatch rather than the decay rate mismatch. We also showed that noise reduction due to the optimized homodyne measurement is remarkable at large cavity driving powers so that it can be increased up to $35~\mathrm{dB}$.
On the other hand, we have demonstrated that in this case, the variational homodyne strategy reduces the signal amplification bandwidth compared to the standard phase detection scheme.
In terms of signal improvement, homodyne detection with an optimized phase angle $\theta_{\rm opt}$ does not provide an improved signal compared to the standard phase detection scheme, in the case of imperfect CQNC conditions and on-resonance driving. In particular, in this case, we showed that signal improvement is $\mathcal{R}=1$ at all frequencies, except in the narrow bandwidth around $\omega_m$ within which $\mathcal{R}<1$. 
Consequently, under these conditions, the variational homodyne CQNC shows simultaneous noise reduction and signal amplification within the frequency bandwidth of $B_{\rm ICQNC}^{\rm hom,on}$ around $\omega_m$, but does not provide an improved signal compared to the standard phase detection scheme.

In the case when the perfect CQNC conditions breaks and the cavity is driven off-resonantly ($\Delta_c\ne0$), noise reduction is again more sensitive to the coupling rate mismatch than to the decay rate mismatch and we have found that noise reduction is enhanced as the parameter $N$ increases, such that for large values of $N$, it can reach $24~\mathrm{dB}$ (see Fig.~\ref{fig13}).
We have also shown that in the case of nonzero detuning, the variational homodyne CQNC scheme shows almost no signal amplification, but provides an improved signal at all frequencies, except in $\Delta\omega_{R<1}$ around $\omega_m$ in which the output signal is weakened. In contrast, the phase detection scheme shows signal amplification within the bandwidth of $B_{\rm ICQNC}^{\rm st, off}$ around $\omega_m$. Consequently, under the imperfect CQNC conditions and off-resonant cavity driving, the variational homodyne detection strategy simultaneously reduces the force noise and provides an improved signal at all frequencies (except in $\Delta\omega_{R<1}$) while it does not show signal amplification. 

We have also demonstrated that the optimized homodyne detection improves the force sensing sensitivity compared to the standard phase detection scheme. The maximum value of sensitivity improvement is $71.60\%$, obtained for the case of perfect CQNC conditions and off-resonant cavity driving, while the sensitivity improvement for imperfect CQNC conditions reduces to less that $10\%$. 

As an outlook, the proposed CQNC assisted by the variational readout can be extended to other optomechanical-like platforms to enhance the performance of the measurement, for example, in a levitated optomechanical-based sensor \cite{bemani2021force}, an ultracoherent optomechanical sensor \cite{Mason2019}, multi-mode optomechanical arrays \cite{Yanay2016,Burgwal2019,Dumont2022,Dumont2022}, a hybrid system \cite{yan2021backaction}, and the ultrasensitive optomechanical gyroscope proposed in \cite{Li2018} which is based on the CQNC method in dual cavity.

\begin{acknowledgements}
The authors would like to thank the referees for their constructive comments, which improved the paper. A.M. wishes to thank the Office of the Vice President for Research of the University of Isfahan and ICQT. H.A. would like to thank the Shahid-Beheshti University. 
A.M. proposed and developed the primary idea of homodyne detection to enhance the sensitivity of CQNC-based quantum sensing, with the help of D.V. All analytical and numerical calculations were performed by H.A. and verified by A.M. 
All authors contributed to the preparation of the paper.
\end{acknowledgements}

\appendix
\section{Spectral density of the added force noise}\label{App.A}
Using Eq.\eqref{eq10}, the power spectral density of the force noise is written as
\begin{align}
&\braket{\hat{F}_{N,\theta}(\omega)\hat{F}_{N,\theta}(-\omega^\prime)}=S_{th}(\omega,\omega^\prime) + S_f(\omega,\omega^\prime) + S_{at}(\omega,\omega^\prime)\notag\\
&\qquad+ S_b(\omega,\omega^\prime)+S_h(\omega,\omega^\prime)+S_{fb}(\omega,\omega^\prime)+S_{fh}(\omega,\omega^\prime)\notag\\
&\qquad+S_{bh}(\omega,\omega^\prime),
\end{align}
where
\begin{equation}\label{eq.A2}
S_{th}(\omega,\omega^\prime)=\braket{\hat{f}(\omega)\hat{f}(-\omega^\prime)},
\end{equation}
\begin{align}
&S_f(\omega,\omega^\prime)= \frac{\kappa}{g^2\omega_m\chi_m(\omega)\chi_m(-\omega^\prime)}\notag\\
&\quad\times\Bigg[\bigg(1-\frac{1}{\kappa\chi_a^\prime(\omega)}\bigg)\bigg(1-\frac{1}{\kappa\chi_a^\prime(-\omega^\prime)}\bigg)\braket{\hat{P}_a^{\rm in}(\omega)\hat{P}_a^{\rm in}(-\omega^\prime)
}\notag\\
&\qquad-\Delta_c\chi_a(-\omega^\prime)\bigg(1-\frac{\kappa}{\kappa\chi_a^\prime(\omega)}\bigg)\braket{\hat{P}_a^{\rm in}(\omega)\hat{X}_a^{\rm in}(-\omega^\prime)}\notag\\
&\qquad-\Delta_c\chi_a(\omega)\bigg(1-\frac{1}{\kappa\chi_a^\prime(-\omega^\prime)}\bigg)\braket{\hat{X}_a^{\rm in}(\omega)\hat{P}_a^{\rm in}(-\omega^\prime) }\notag\\
&\qquad +\Delta_c^2\chi_a(\omega)\chi_a(-\omega^\prime)\braket{\hat{X}_a^{\rm in}(\omega)\hat{X}_a^{\rm in}(-\omega^\prime)} \Bigg],\label{eq.A3}
\end{align}
\begin{align}
&S_{at}(\omega,\omega^\prime)=\frac{G^2\Gamma\chi_d(\omega)\chi_d(-\omega^\prime)}{g^2\gamma_m\chi_m(\omega)\chi_m(-\omega^\prime)}\Bigg[\braket{\hat{P}_d^{\rm in}(\omega)\hat{P}_d^{\rm in}(-\omega^\prime)}\notag\\
&\quad -\frac{\Gamma/2 - i\omega^\prime}{\omega_m}\braket{\hat{P}_d^{\rm in}(\omega)\hat{X}_d^{\rm in}(-\omega^\prime)}\notag\\
&\quad- \frac{\Gamma/2 + i\omega}{\omega_m}\braket{\hat{X}_d^{\rm in}(\omega)\hat{P}_d^{\rm in}(-\omega^\prime)}\notag\\
&\quad +\frac{(\Gamma/2 + i\omega)(\Gamma/2 - i\omega^\prime)}{\omega_m^2}\braket{\hat{X}_a^{\rm in}(\omega)\hat{X}_a^{\rm in}(-\omega^\prime)} \Bigg],\label{eq.A4}
\end{align}
\begin{align}
&S_b(\omega,\omega^\prime)=\frac{\kappa}{g^2\gamma_m}\frac{g^2\chi_m(\omega)+G^2\chi_d(\omega)}{\chi_m(\omega)}\notag\\
&\quad\times\frac{g^2\chi_m(-\omega^\prime)+G^2\chi_d(-\omega^\prime)}{\chi_m(-\omega^\prime)}\braket{\hat{X}_a^{\rm in}(\omega)\hat{X}_a^{\rm in}(-\omega^\prime)}, \label{eq.A5}
\end{align}
\begin{align}
&S_h(\omega,\omega^\prime)=\frac{\mathcal{B}^2}{g^2\kappa\chi_m(\omega)\chi_m(-\omega^\prime)\chi_a^\prime(\omega)\chi_a^\prime(-\omega^\prime)}\notag\\
&\quad\times\Big\{ \Delta_c^2\chi_a(\omega)\chi_a(-\omega^\prime)\braket{\hat{P}_a^{\rm in}(\omega)\hat{P}_a^{\rm in}(-\omega^\prime)}\notag\\
&\qquad+\Delta_c\chi_a(\omega)[\kappa\chi_a(-\omega^\prime)-1]\braket{\hat{P}_a^{\rm in}(\omega)\hat{X}_a^{\rm in}(-\omega^\prime)}\notag\\
&\qquad+\Delta_c\chi_a(-\omega^\prime)[\kappa\chi_a(\omega)-1]\braket{\hat{X}_a^{\rm in}(\omega)\hat{P}_a^{\rm in}(-\omega^\prime)}\notag\\
&\qquad+[\kappa\chi_a(\omega)-1][\kappa\chi_a(-\omega^\prime)-1]\braket{\hat{X}_a^{\rm in}(\omega)\hat{X}_a^{\rm in}(-\omega^\prime)} \Big\},\label{eq.A6}
\end{align}
\begin{align}
&S_{fh}(\omega,\omega^\prime)=-\frac{\mathcal{B}}{g^2\gamma_m\chi_m(\omega)\chi_m(-\omega^\prime)\chi_a^\prime(-\omega^\prime)}\notag\\
&\quad\times\Bigg[\Delta_c\chi_a(-\omega^\prime)\bigg(1-\frac{1}{\kappa\chi_a^\prime(\omega)}\bigg)\braket{\hat{P}_a^{\rm in}(\omega)\hat{P}_a^{\rm in}(-\omega^\prime)}\notag\\
&\qquad+[\kappa\chi_a(-\omega^\prime)-1]\bigg(1-\frac{1}{\kappa\chi_a^\prime(\omega)}\bigg)\braket{\hat{P}_a^{\rm in}(\omega)\hat{X}_a^{\rm in}(-\omega^\prime)}\notag\\
&\qquad -\Delta_c^2\chi_a(\omega)\chi_a(-\omega^\prime)\braket{\hat{X}_a^{\rm in}(\omega)\hat{P}_a^{\rm in}(-\omega^\prime)}\notag\\
&\qquad - \Delta_c\chi_a(\omega)[\kappa\chi_a(-\omega^\prime)-1]\braket{\hat{X}_a^{\rm in}(\omega)\hat{X}_a^{\rm in}(-\omega^\prime)} \Bigg]\notag\\
&\quad -\frac{\mathcal{B}}{g^2\gamma_m\chi_m(\omega)\chi_m(-\omega^\prime)\chi_a^\prime(\omega)}\notag\\
&\qquad \times\Bigg[\Delta_c\chi_a(\omega)\bigg(1-\frac{1}{\kappa\chi_a^\prime(-\omega^\prime)}\bigg)\braket{\hat{P}_a^{\rm in}(\omega)\hat{P}_a^{\rm in}(-\omega^\prime)}\notag\\
&\qquad\quad -\Delta_c^2\chi_a(\omega)\chi_a(-\omega^\prime)\braket{\hat{P}_a^{\rm in}(\omega)\hat{X}_a^{\rm in}(-\omega^\prime)}\notag\\
&\qquad\quad +[\kappa\chi_a(\omega)-1]\bigg(1-\frac{1}{\kappa\chi_a^\prime(-\omega^\prime)}\bigg)\braket{\hat{X}_a^{\rm in}(\omega)\hat{P}_a^{\rm in}(-\omega^\prime)}\notag\\
&\qquad\quad -\Delta_c\chi_a(-\omega^\prime)[\kappa\chi_a(\omega)-1]\braket{\hat{X}_a^{\rm in}(\omega)\hat{X}_a^{\rm in}(-\omega^\prime)} \Bigg],\label{eq.A7}
\end{align}
\begin{align}
&S_{fb}(\omega,\omega^\prime)=\frac{\kappa}{g^2\gamma_m}\frac{g^2\chi_m(-\omega^\prime)+G^2\chi_d(-\omega^\prime)}{\chi_m(\omega)\chi_m(-\omega^\prime)}\chi_a(-\omega^\prime)\notag\\
&\quad\times\Bigg[\big(1-\frac{1}{\kappa\chi_a^\prime(\omega)}\big)\braket{\hat{P}_a^{\rm in}(\omega)\hat{X}_a^{\rm in}(-\omega^\prime)}\notag\\
&\qquad-\Delta_c\chi_a(\omega)\braket{\hat{X}_a^{\rm in}(\omega)\hat{X}_a^{\rm in}(-\omega^\prime)} \Bigg]\notag\\
&+ \frac{\kappa}{g^2\gamma_m}\frac{g^2\chi_m(\omega)+G^2\chi_d(\omega)}{\chi_m(\omega)\chi_m(-\omega^\prime)}\chi_a(\omega)\notag\\
&\quad\times\Bigg[\big(1-\frac{1}{\kappa\chi_a^\prime(-\omega^\prime)}\big)\braket{\hat{X}_a^{\rm in}(\omega)\hat{P}_a^{\rm in}(-\omega^\prime)}\notag\\
&\qquad-\Delta_c\chi_a(-\omega^\prime)\braket{\hat{X}_a^{\rm in}(\omega)\hat{X}_a^{\rm in}(-\omega^\prime)} \Bigg], \label{eq.A8}
\end{align}
\begin{align}
&S_{bh}(\omega,\omega^\prime)=-\frac{\mathcal{B}}{g^2\gamma_m}\frac{g^2\chi_m(\omega)+G^2\chi_d(\omega)}{\chi_m(\omega)\chi_m(-\omega^\prime)\chi_a^\prime(-\omega^\prime)}\chi_a(\omega)\notag\\
&\quad\times\Big\{\Delta_c\chi_a(-\omega^\prime)\braket{\hat{X}_a^{\rm in}(\omega)\hat{P}_a^{\rm in}(-\omega^\prime)}\notag\\
&\qquad+[\kappa\chi_a(-\omega^\prime)-1]\braket{\hat{X}_a^{\rm in}(\omega)\hat{P}_a^{\rm in}(-\omega^\prime)}\Big\}\notag\\
&\quad-\frac{\mathcal{B}}{g^2\gamma_m}\frac{g^2\chi_m(-\omega^\prime)+G^2\chi_d(-\omega^\prime)}{\chi_m(\omega)\chi_m(-\omega^\prime)\chi_a^\prime(\omega)}\chi_a(-\omega^\prime)\notag\\
&\qquad\times\Big\{\Delta_c\chi_a(\omega)\braket{\hat{P}_a^{\rm in}(\omega)\hat{X}_a^{\rm in}(-\omega^\prime)}\notag\\
&\quad\qquad +[\kappa\chi_a(\omega)-1]\braket{\hat{X}_a^{\rm in}(\omega)\hat{X}_a^{\rm in}(-\omega^\prime)} \Big\}.\label{eq.A9}
\end{align}
The correlation functions appearing in Eqs.~\eqref{eq.A2}--\eqref{eq.A9} in the Fourier space are given by
\begin{subequations}
	\begin{align}
	\braket{\hat{f}(\omega)\hat{f}(-\omega^\prime)}&=(\bar{n}_m+\frac{1}{2})\delta(\omega-\omega^\prime)\simeq\frac{k_BT}{\hbar\omega_m}\delta(\omega-\omega^\prime),\\
	\braket{\hat{X}_a^{\rm in}(\omega)\hat{X}_a^{\rm in}(-\omega^\prime)}&=\bigg(N+\frac{1}{2}+\mathrm{Re}M\bigg)\delta(\omega-\omega^\prime),\\
	\braket{\hat{P}_a^{\rm in}(\omega)\hat{P}_a^{\rm in}(-\omega^\prime)}&=\bigg(N+\frac{1}{2}-\mathrm{Re}M\bigg)\delta(\omega-\omega^\prime),\\
	\braket{\hat{X}_a^{\rm in}(\omega)\hat{P}_a^{\rm in}(-\omega^\prime)}&=\frac{i}{2}(1-2i\mathrm{Im}M)\delta(\omega-\omega^\prime),\\
	\braket{\hat{P}_a^{\rm in}(\omega)\hat{X}_a^{\rm in}(-\omega^\prime)}&=-\frac{i}{2}(1+2i\mathrm{Im}M)\delta(\omega-\omega^\prime),\\
	\braket{\hat{X}_d^{\rm in}(\omega)\hat{X}_d^{\rm in}(-\omega^\prime)}&=\braket{\hat{P}_d^{\rm in}(\omega)\hat{P}_d^{\rm in}(-\omega^\prime)}=\frac{1}{2}\delta(\omega-\omega^\prime),\\
	\braket{\hat{P}_d^{\rm in}(\omega)\hat{X}_d^{\rm in}(-\omega^\prime)}&=-\braket{\hat{X}_d^{\rm in}(\omega)\hat{P}_d^{\rm in}(-\omega^\prime)}=\frac{i}{2}\delta(\omega-\omega^\prime).
	\end{align}
\end{subequations}
Using these expressions, one finally gets the generalized spectral density of added force noise, Eq.~\eqref{eq12}.

\section{Estimation of the noise reduction in the case of imperfect CQNC and $G=g$}\label{app.B}
According to Eq.~\eqref{eq32}, the amount of noise reduction $\Delta S$ due to homodyne detection with optimum phase angle $\theta_{\rm opt}$ is proportional to
\begin{equation}
\Bigg[\mathrm{Re}\big(\frac{1+(G/g)^2R(\omega)}{\chi_m(\omega)}\big)\Bigg]^2|\chi_m(\omega)|^2.
\end{equation}
In the Markovian limit, $\kappa\gg\omega_m$, we keep only the zeroth order of $\omega/\kappa$ and we therefore have
\begin{align}
\mathrm{Re}\big(\frac{1+(G/g)^2R(\omega)}{\chi_m(\omega)}\big)\simeq&\omega_m\bigg(1-\frac{\omega^2}{\omega_m^2}\bigg)^2\bigg(1+\frac{G^2}{g^2}R_r \bigg)\notag\\
&+ \frac{\omega\gamma_m}{\omega_m}\frac{G^2}{g^2}R_i,
\end{align}
where
\begin{subequations}
	\begin{align}
	R_r &= -1-\frac{\omega^2\Gamma(\gamma_m-\Gamma)}{(\omega_m^2-\omega^2)^2+\omega^2\Gamma^2},\\
	R_i&=\frac{\omega(\gamma_m-\Gamma)(\omega_m^2-\omega^2)}{(\omega_m^2-\omega^2)^2+\omega^2\Gamma^2}.
	\end{align}
\end{subequations}
By considering  the same parameters as those used in Fig.~\ref{fig7}, we conclude that
\begin{equation}
\Bigg[\mathrm{Re}\big(\frac{1+(G/g)^2R(\omega)}{\chi_m(\omega)}\big)\Bigg]^2|\chi_m(\omega)|^2\bigg|_{\omega=\omega_m+4\gamma_m}\simeq\bigg(1-\frac{\Gamma}{\gamma_m}\bigg)^2.
\end{equation}
This result shows that in the case of imperfect CQNC and $G=g$, the amount of noise reduction is
\begin{equation}
\Delta S\big|_{\omega=\omega_m+4\gamma_m}\propto\bigg(1-\frac{\Gamma}{\gamma_m}\bigg)^2.
\end{equation}

\section{Estimation of the optical susceptibility $\chi_a(\omega)$ \label{app. C}}
\begin{figure}
	\begin{center}
		\includegraphics[width=9cm]{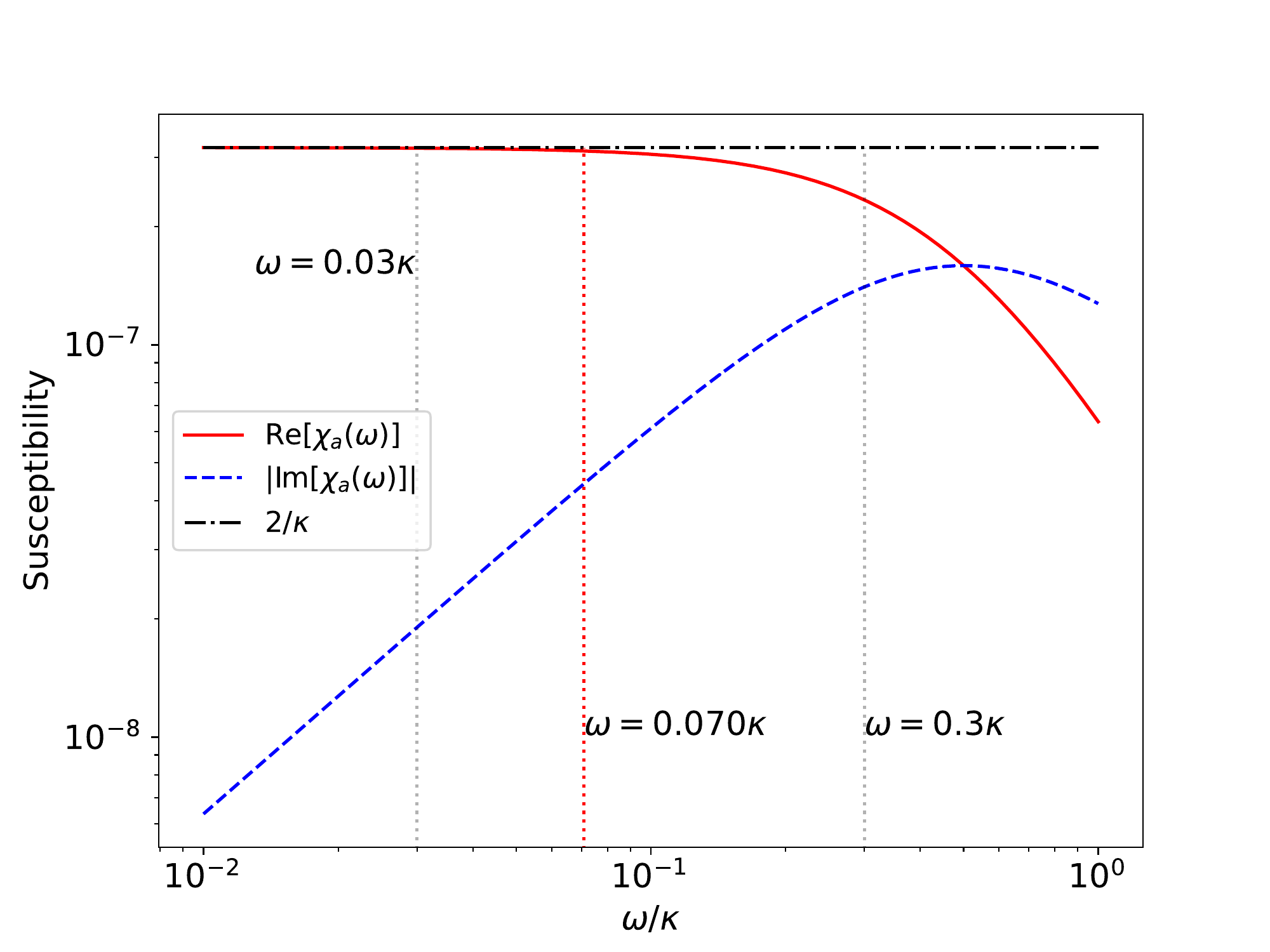}
	\end{center}
	\vspace*{-5mm}
	\caption{(Color online) Optical susceptibility $\chi_a(\omega)$ versus the scaled frequency $\omega/\kappa$. Different curves correspond to $\mathrm{Re}[\chi_a(\omega)]$ (red solid line), $\mathrm{Im}[\chi_a(\omega)]$ (blue dashed line), and $\chi_a(\omega)\sim 2/\kappa$ (black dash-dotted line).  }
	\label{figC1}
\end{figure}
According to Eq.~\eqref{eq7a}, the optical susceptibility $\chi_a(\omega)$ can be written as
\begin{equation} \label{eqC1}
	\chi_a(\omega)=\mathrm{Re}[\chi_a(\omega)]+i\mathrm{Im}[\chi_a(\omega)],
\end{equation}
with
\begin{align}
	\mathrm{Re}[\chi_a(\omega)]&=\frac{\kappa/2}{\kappa^2/4+\omega^2}\\
	\mathrm{Im}[\chi_a(\omega)]&=-\frac{\omega}{\kappa^2/4+\omega^2}.
\end{align}
From these equations one can see that by decreasing the ratio $\omega/\kappa$, the imaginary part of $\chi_a(\omega)$ decreases. In Fig.~\ref{figC1} we illustrate the optical susceptibility versus the scaled frequency $\omega/\kappa$. This figure shows that if the ratio $\omega/\kappa$ is less than 0.070, approximating the optical susceptibility as $\chi_a\sim 2/\kappa$ introduces lower than $1\%$ error in the value of $|\chi_a(\omega)|$, while for $\omega/\kappa=0.03$ the error is $0.1\%$. Throughout the paper, we choose the frequency $\omega$ of the order of the mechanical resonance frequency $\omega_m$ and we choose the parameters so that $\omega_m/\kappa=0.03$ to satisfy the resolved sideband regime. As a result, approximating the optical susceptibility as $\chi_a(\omega)\sim 2/\kappa$ does not introduce a substantial error to the calculations. However, for $ \omega_m/\kappa=0.3 $ it causes $ 17\% $ error.

In the case of perfect CQNC conditions and off-resonant cavity driving, the noise reduction advantage given by Eq.~(\ref{eqold30}) is proportional to $ (\omega_m/\kappa)^{-1} $ and therefore the smaller the value of $ \omega_m/\kappa $, the greater the noise reduction advantage.

In the case of imperfect CQNC conditions, a completely different behavior is obtained. Based on Eq.~(\ref{eq32}), when the imperfect CQNC conditions hold and the cavity is driven resonantly, the noise reduction advantage is proportional to $ \omega_m/\kappa $. Therefore, lowering the ratio $ \omega_m/\kappa $ leads to the noise reduction advantage which shows the advantage of the variational homodyne detection over the phase detection scheme versus the frequency. This means that the advantage of the variational homodyne detection disappears for small values of $ \omega_m/\kappa $, i.e., the strong resolved-sideband regime.

\section{Estimation of the squeezing parameter N}\label{app.D}
The quantum description of the electromagnetic field predicts a class of states for which the variances in the field quadrature operators $\hat{X}_1=(\hat{a}^\dagger + \hat{a})/\sqrt{2}$ and $\hat{X}_2=(\hat{a}^\dagger-\hat{a})/i\sqrt{2}$, with $\hat{a}$ ($\hat{a}^\dagger$) the annihilation (creation) operator of the light field, are limited by
\begin{equation}\label{eq.D1}
	\sigma_{X_1}\sigma_{X_2}=1/4,
\end{equation} 										
in which $\sigma_{X_i}=\sqrt{\braket{\hat{X}_i^2}-\braket{\hat{X}_i}^2}$, with $i=1, 2$. Coherent light, for which the variances of the two quadratures are equal, belongs to this class. Reducing the uncertainty in one quadrature below this state refers to squeezing. Equation \eqref{eq.D1} dictates that squeezing of one quadrature must increase the uncertainty in the orthogonal quadrature, an effect known as antisqueezing. A detailed description of quantum squeezing can be found in standard textbooks of quantum optics \cite{Sculy1997quantumoptics}. In the following, we introduce a brief review on such states.

Theoretically, the squeezed state of light is obtained by operating the squeezing operator on the vacuum state of light.  The squeezing operator is defined by
\begin{equation}\label{D2}
	\hat{S}=\exp\bigg[ \frac{1}{2}\xi^* \hat{a}^2 - \frac{1}{2}\xi (\hat{a}^\dagger)^2 \bigg],
\end{equation}
 where $\xi=r e^{i\Phi}$, with $r$ and $\Phi$ the squeezing strength and phase, respectively. It is straightforward to show that the variances of the squeezed state of light are given by $\sigma_{X_1}^2=e^{-2r}/4$ and $\sigma_{X_2}^2=e^{2r}/4$. It is obvious that the squeezing is characterized by $e^{-2r}$. Consequently, it is useful to define an experimental quantity as the squeezing level which is defined as $V_-=e^{-2r}$ \cite{Zuo2022}. As mentioned in \cite{motazedifard2016force}, one can represent the squeezing parameter $N$ in terms of the squeezing strength $r$ as $N=\sinh^2(r)$. Combining these expressions, one can rewrite the squeezing parameter in terms of the squeezing level as
\begin{equation}\label{D3}
	N=\frac{1}{4}(V_-^{-1/2}-V_-^{1/2})^2.
\end{equation}

Experimental realization of squeezed states is based on the nonlinear optical phenomena \cite{Sculy1997quantumoptics}. The most efficient method of squeezed state generation utilizes a subthreshold optical parametric oscillator (OPO), in which a nonlinear crystal, for example $\beta$ barium borate (BBO), periodically poled lithium niobate (PPLN), or periodically poled titanyl phosphate (PPKTP) (for more details see Refs.~\cite{shySPDC,aliBellBBO,aliDNA,FrequencyDependentSqueezing,aliBellPPKTP,boyd} and references therein), is used as a nonlinear medium in the OPO. By utilizing the PPKTP crystal, Takeno\textit{ et. al.} obtained a squeezing level of $-9~\mathrm{dB}$ \cite{Takeno2007}, which is equivalent to the squeezing parameter of $N=3.0349$, while in 2022, Dwyer \textit{et al.} demonstrated a squeezing level of approximately $-20~\mathrm{dB}$ \cite{Dwyer2022}, which is equivalent to the squeezing parameter of $N\simeq24.37$ and the highest squeezing level reported to date. That is why we have considered $ N_{\rm max} = 25 $ in the present paper.


\begin{thebibliography}{60}
	\setlength{\baselineskip}{0.2 \baselineskip}
	

	\bibitem{stannigel2012optomechanical} K. Stannigel, P. Komar, S. J. M. Habraken, S. D. Bennett, M. D. Lukin, P. Zoller, and P. Rabl, \href{https://doi.org/10.1103/PhysRevLett.109.013603}{Phys. Rev. Lett. \textbf{109}, 013603 (2012)}. 
	
	\bibitem{stannigel2011optomechanical}K. Stannigel, P. Rabl, A. S. Sørensen, M. D. Lukin, and P. Zoller,
	\href{https://doi.org/10.1103/PhysRevA.84.042341}{Phys. Rev. A \textbf{84}, 042341 (2011)}. 
	
	\bibitem{stannigel2010optomechanical}K. Stannigel, P. Rabl, A. S. Sørensen, P. Zoller, and M. D. Lukin, \href{https://doi.org/10.1103/PhysRevLett.105.220501}{Phys. Rev.
Lett. \textbf{105}, 220501 (2010)}. 
	
	\bibitem{rogers2014hybrid} B. Rogers, N. Lo Gullo, G. De Chiara, G. M. Palma, and M. Paternostro,
	\href{https://doi.org/10.2478/qmetro-2014-0002}{ Quantum Meas. Quantum Metrol. \textbf{2}, 11 (2014)}. 
	
	\bibitem{sete2015high} E. A. Sete and H. Eleuch, \href{https://doi.org/10.1103/PhysRevA.91.032309}{Phys. Rev. A \textbf{91}, 032309 (2015)}. 
	
	\bibitem{wallucks2020quantum}A. Wallucks, I. Marinković, B. Hensen, R. Stockill, and S. Gröblacher,
	\href{https://doi.org/10.1038/s41567-020-0891-z}{Nat. Phys. \textbf{16}, 772 (2020)}.
	
	\bibitem{andrews2014bidirectional}R. W. Andrews, R. W. Peterson, T. P. Purdy, K. Cicak, R. W. Simmonds, C. A. Regal, and K. W. Lehnert,
\href{https://doi.org/10.1038/nphys2911}{Nat. Phys. \textbf{10}, 321 (2014).} 
	
	\bibitem{forsch2020microwave} M. Forsch, R. Stockill, A. Wallucks, I. Marinković, C. Gärtner, R. A. Norte, F. van Otten, A. Fiore, K. Srinivasan, and S. Gröblacher,
	\href{https://doi.org/10.1038/s41567-019-0673-7}{Nat. Phys. \textbf{16}, 69 (2020).} 
	
	\bibitem{jiang2020efficient} W. Jiang, C. J. Sarabalis, Y. D. Dahmani, R. N. Patel, F. M. Mayor, T. P. McKenna, R. V. Laer, and A. H. Safavi-Naeini,
\href{https://doi.org/10.1038/s41467-020-14863-3}{Nature Communications \textbf{11}, 1166 (2020).} 
	
	\bibitem{lauk2020perspectives}N. Lauk, N. Sinclair, S. Barzanjeh, J. P. Covey, M. Saffman, M. Spiropulu, and C. Simon,
	\href{https://doi.org/10.1088/2058-9565/ab788a}{Quantum Science and Technology \textbf{5}, 020501 (2020)}. 
	
	\bibitem{arnold2020converting} G. Arnold, M. Wulf, S. Barzanjeh, E. S. Redchenko, A. Rueda, W. J. Hease, F. Hassani, and J. M. Fink,
\href{https://doi.org/10.1038/s41467-020-18269-z}{Nature communications \textbf{11}, 4460 (2020).} 
	
	\bibitem{barzanjeh2017mechanical} S. Barzanjeh, M. Wulf, M. Peruzzo, M. Kalaee, P. B. Dieterle, O. Painter, and J. M. Fink,
\href{https://doi.org/10.1038/s41467-017-01304-x}{Nature communications \textbf{8}, 953 (2017)}. 
	
	\bibitem{shen2018reconfigurable} Z. Shen, Y.-L. Zhang, Y. Chen, F.-W. Sun, X.-B. Zou, G.-C. Guo, C.-L. Zou, and C.-H. Dong,
\href{https://doi.org/10.1038/s41467-018-04187-8}{Nature communications \textbf{9}, 1797 (2018)}. 
	
	
	\bibitem{foroudcrystalentanglement} F. Bemani, R. Roknizadeh, A. Motazedifard, M. H. Naderi, and D. Vitali, 
	\href{https://doi.org/10.1103/PhysRevA.99.063814}{ Phys. Rev. A \textbf{99}, 063814 (2019)}. 
	
	\bibitem{polzikDistantEntanglementOMS2020} R. A. Thomas, M. Parniak, C. Ostfeldt, C. B. Moller, C. Baerentsen, Y. Tsaturyan, A. Schliesser, J. Appel, E. Zeuthen, and E.
S. Polzik,
	\href{https://doi.org/10.1038/s41567-020-1031-5}{Nat. Phys. \textbf{17}, 228 (2021).} 
	
	\bibitem{foroudsynch} F. Bemani, A. Motazedifard, R. Roknizadeh, M. H. Naderi, and D. Vitali,
	\href{https://journals.aps.org/pra/abstract/10.1103/PhysRevA.96.023805}{Phys. Rev. A \textbf{96}, 023805 (2017)}. 
	
	\bibitem{barzanjehentanglement} S. Barzanjeh, E. S. Redchenko, M. Peruzzo, M. Wulf, D. P. Lewis, G. Arnold, and J. M. Fink,
	\href{https://doi.org/10.1038/s41586-019-1320-2}{Nature \textbf{570}, 480 (2019)}.
	
	\bibitem{dalafiQOC} A. Dalafi, M. H. Naderi, and A. Motazedifard,
	\href{https://doi.org/10.1103/PhysRevA.97.043619}{Phys. Rev. A \textbf{97}, 043619 (2018)}. 
	
	
	\bibitem{giovanniCQNC2020} A. Di Giovanni, M. Brunelli, and M. G. Genoni,
	\href{https://link.aps.org/doi/10.1103/PhysRevA.103.022614}{Phys. Rev. A \textbf{103}, 022614 (2021)}. 
	
	\bibitem{aliDCEsqueezing} A. Motazedifard, A. Dalafi, M. H. Naderi, and R. Roknizadeh,
	\href{https://doi.org/10.1016/j.aop.2019.03.019}{Ann. Phys. (NY) \textbf{405}, 202 (2019)}.
	
	
	\bibitem{teleportation} J. Li, A. Wallucks, R. Benevides, N. Fiaschi, B. Hensen, T. P. M. Alegre, and S. Gr{\"o}blacher,
	\href{https://doi.org/10.1103/PhysRevA.102.032402}{Phys. Rev. A \textbf{102}, 032402 (2020)}.
	
	
	\bibitem{Pikovski2012} I. Pikovski, M. R. Vanner, M. Aspelmeyer, M. Kim, and C. Brukner,
	\href{https://doi.org/10.1038/nphys2262}{Nat. Phys. \textbf{8}, 393 (2012)}.
	
	
	\bibitem{PhysRevLett.116.070405} V. C. Vivoli, T. Barnea, C. Galland, and N. Sangouard,
	\href{https://link.aps.org/doi/10.1103/PhysRevLett.116.070405}{Phys. Rev. Lett. \textbf{116}, 070405 (2016)}. 
	
	\bibitem{PhysRevLett.107.020405} O. Romero-Isart, A. C. Pflanzer, F. Blaser, R. Kaltenbaek, N. Kiesel, M. Aspelmeyer, and J. I. Cirac,
	\href{https://link.aps.org/doi/10.1103/PhysRevLett.107.020405}{Phys. Rev. Lett. \textbf{107}, 020405 (2011)}.
	
	\bibitem{optomechanicalBelltest1} I. Marinković, A. Wallucks, R. Riedinger, S. Hong, M. Aspelmeyer, and S. Gr{\"o}blacher, 
	\href{https://doi.org/10.1103/PhysRevLett.121.220404}{Phys. Rev. Lett. \textbf{121}, 220404 (2018)}.
	
	
	\bibitem{aliDCE2} A. Motazedifard, M. H. Naderi, and R. Roknizadeh, 
	\href{https://doi.org/10.1364/JOSAB.34.000642}{J. Opt. Soc. Am. B \textbf{34}, 642 (2017)}.
	
	
	\bibitem{aliDCE3} A. Motazedifard, A. Dalafi, M. H. Naderi, and R. Roknizadeh, \href{https://doi.org/10.1016/j.aop.2018.07.013}{Ann. Phys. (NY) \textbf{396}, 202 (2018)}.
	
	\bibitem{NoriDCE1} O. Di Stefano, A. Settineri, V. Macrì, A. Ridolfo, R. Stassi, A. F. Kockum, S. Savasta, and F. Nori, 
	\href{https://doi.org/10.1103/PhysRevLett.122.030402}{Phys. Rev. Lett. 122, 030402 (2019)}.
	
	\bibitem{aliGreen2021} A Motazedifard, A. Dalafi, M. H. Naderi, 
	\href{https://doi.org/10.1088/1751-8121/abf3e9}{J. Phys. A: Math. Theor. \textbf{54} (21), 215301 (2021)}.
	
	\bibitem{Heforce2022} Q. He, F. Badshah, Y. Song, L. Wang, E. Liang, and S.L. Su,
	\href{https://link.aps.org/doi/10.1103/PhysRevA.105.013503}{Phys. Rev. A \textbf{105}, 013503 (2022)}. 
	
	
	\bibitem{braginsky1995quantum}V. B. Braginsky, F. Y. Khalili, and S. Thorne, \textit{Quantum Measurement}, (Cambridge University Press,Cambridge,1992). 
	
	\bibitem{virgoBAexperiment2020} F. Acernese \textit{et al}. (The Virgo Collaboration),
	\href{https://doi.org/10.1103/PhysRevLett.125.131101}{Phys. Rev. Lett. \textbf{125}, 131101 (2020)}. 
	
	
	\bibitem{aspelmeyer2014cavity}M. Aspelmeyer, T. J. Kippenberg, and F. Marquardt,
	\href{https://doi.org/10.1103/RevModPhys.86.1391}{Rev. Mod. Phys. \textbf{86}, 1391 (2014)}. 
	
	\bibitem{meystre2013short}P. Meystre,
	\href{https://doi.org/10.1002/andp.201200226}{Ann. Phys. (Berlin) \textbf{525}, 215 (2013)}. 
	
	\bibitem{teufel2009nanomechanical} J. D. Teufel, T. Donner, M. A. Castellanos-Beltran, J. W. Harlow, and K. W. Lehnert,
\href{https://doi.org/10.1038/nnano.2009.343}{Nature Nanotech \textbf{4}, 820 (2009)}. 
	
	
	\bibitem{huang2017robust}S. Huang and G. S. Agarwal, \href{https://doi.org/10.1103/PhysRevA.95.023844}{Phys. Rev. A \textbf{95}, 023844 (2017)}. 
	
	
	\bibitem{chao2022backaction}S.-L. Chao, D.-W. Wang, Z. Yang, C.-S. Zhao, R. Peng, and L. Zhou \href{https://doi.org/10.1002/andp.202100421}{ Ann. Phys. (Berlin) \textbf{534},2100421 (2022)}. 
	
	\bibitem{fani2020back} M. Fani and A. Dalafi, \href{https://doi.org/10.1364/JOSAB.386227}{J. Opt. Soc. Am. B \textbf{37}, 1263 (2020)}. 
	
	\bibitem{clerk2008back} A. A. Clerk, F. Marquardt, and K. Jacobs,
	\href{https://doi.org/10.1088/1367-2630/10/9/095010}{New J. Phys. \textbf{10}, 095010 (2008)}. 
	
	
	
	\bibitem{sillanpaa2020ForceFree} L. M. de Lépinay, C. F. Ockeloen-Korppi, M. J. Woolley,  and M. A. Sillanpää, \href{https://www.science.org/doi/abs/10.1126/science.abf5389}{Science \textbf{372}, 6542 (2021)}. 
	
	
	\bibitem{Kampel2017improve}N. S. Kampel, R. W. Peterson, R. Fischer, P.-L. Yu,  K. Cicak, R. W. Simmonds, K. W. Lehnert, and C. A. Regal,
    \href{https://link.aps.org/doi/10.1103/PhysRevX.7.021008}{Phys. Rev. X \textbf{7}, 021008 (2017)}
	
	
	
	\bibitem{bemani2021force} F. Bemani, O. Černotík, L. Ruppert, D. Vitali, and R. Filip,
	\href{https://arxiv.org/abs/2106.11199v1}{Phys. Rev. Applied \textbf{17}, 034020 (2022)}. 
	
	\bibitem{tsang2010coherent}M. Tsang and C. M. Caves, \href{https://doi.org/10.1103/PhysRevLett.105.123601}{Phys. Rev. Lett. \textbf{105}, 123601 (2010)}. 
	
	\bibitem{gong2021weak}B. Gong, D. Dong, and W. Cui, \href{https://doi.org/10.1088/1751-8121/abe888}{J. Phys. A: Math. Theor. \textbf{54}, 165301 (2021)}. 
	
	\bibitem{lee2020squeezed}C. W. Lee, J. H. Lee, and H. Seok, \href{https://doi.org/10.1038/s41598-020-74629-1}{Sci. Rep. \textbf{10}, 17496 (2020)}. 

	
	\bibitem{aliDCEforcesenning} A. Motazedifard, A. Dalafi, F. Bemani, and M. H. Naderi,
	\href{https://doi.org/10.1103/PhysRevA.100.023815}{Phys. Rev. A \textbf{100}, 023815 (2019)}. 
	
	\bibitem{mehryMagneticsensing2020} M. S. Ebrahimi, A. Motazedifard, and M. Bagheri Harouni,
	\href{https://link.aps.org/doi/10.1103/PhysRevA.103.062605}{Phys. Rev. A \textbf{103}, 062605 (2021)}.
	
	\bibitem{Jong2022cooperativity}M. de Jong, J. Li, C. Gärtner, R. Norte, and S. Gröblacher, 
	\href{https://doi.org/10.1364/OPTICA.446434}{Optica \textbf{9}, 170-176 (2022)}.
	
	
	\bibitem{Yanay2016} Y. Yanay, J. C. Sankey, and A. A. Clerk,
	\href{https://link.aps.org/doi/10.1103/PhysRevA.93.063809}{Phys. Rev. A \textbf{93}, 063809 (2016)}. 
	
	\bibitem{Burgwal2019} R. Burgwal, J. del Pino, and E. Verhagen
	\href{http://dx.doi.org/10.1088/1367-2630/abc1c8}{New J. Phys. \textbf{22}, 113006 (2020)}. 
	
	\bibitem{Dumont2022} V. Dumont, H.-K. Lau, A. A. Clerk, and J. C. Sankey,
	\href{https://doi.org/10.48550/arXiv.2203.00631}{arXiv:2203.00631 [quant-ph] (2022)}.
	
	\bibitem{Paraiso2015} T. K. Para{\"i}so, M. Kalaee, L. Zang, H. Pfeifer, F. Marquardt, and O. Painter, \href{https://link.aps.org/doi/10.1103/PhysRevX.5.041024}{Phys. Rev. X \textbf{5}, 041024 (2015)}. 
	
	
	\bibitem{motazedifard2021ultraprecision}A. Motazedifard, A. Dalafi, and M. H. Naderi,
	\href{https://doi.org/10.1116/5.0035952}{AVS Quantum Sci. \textbf{3}, 024701 (2021)}. 
	
	\bibitem{motazedifard2016force}A. Motazedifard, F. Bemani, M. H. Naderi, R. Roknizadeh, and D. Vitali,
	\href{https://doi.org/10.1088/1367-2630/18/7/073040}{New J. Phys. \textbf{18}, 073040 (2016)}. 
	
	\bibitem{bariani2015atom}F. Bariani, H. Seok, S. Singh, M. Vengalattore, and P. Meystre,
	\href{https://doi.org/10.1103/PhysRevA.92.043817}{Phys. Rev. A \textbf{92}, 043817  (2015)}. 
	
	
	\bibitem{cQNCNatureexp} C. B. Møller, R. A. Thomas, G. Vasilakis, E. Zeuthen, Y. Tsaturyan, M. Balabas, K. Jensen, A. Schliesser, K. Hammerer, and E. S. Polzik,
	\href{https://doi.org/10.1038/nature22980}{Nature (London) \textbf{547}, 191 (2017)}. 
	
    \bibitem{Kimble2001}H. J. Kimble, Y. Levin, A. B. Matsko, K. S. Thorne, and S. P. Vyatchanin, \href{https://link.aps.org/doi/10.1103/PhysRevD.65.022002}{Phys. Rev. D \textbf{65}, 022002 (2001)} . 


	\bibitem{Mason2019} D. Mason, J. Chen, M. Rossi, Y. Tsaturyan, and A. Schliesser, \href{https://doi.org/10.1038/s41567-019-0533-5}{Nat. Phys. \textbf{15}, 745–749 (2019)} 

	
	\bibitem{giovannetti2001phase}V. Giovannetti and D. Vitali, \href{https://doi.org/10.1103/PhysRevA.63.023812}{Phys. Rev. A \textbf{63}, 023812 (2001)}. 
	

	\bibitem{yan2021backaction}J.-S. Yan and J. Jing, \href{https://doi.org/10.1002/andp.202100119}{Ann. Phys. (Berlin) \textbf{533}, 2100119 (2021)}. 
		
		
		
		
	\bibitem{Li2018}K. Li, S. Davuluri, and Y. Li, \href{https://doi.org/10.1007/s11433-018-9189-6}{Sci. China Phys. Mech. Astron. \textbf{61}, 90311 (2018)}. 
	
	\bibitem{Sculy1997quantumoptics}M. O. Scully,  M. S. Zubairy, \textit{Quantum Optics} (Cambridge University Press, Cambridge,1997). 
	
	\bibitem{Zuo2022} G. Zuo, Y. Zhang, J. Li, S. Zhu, Y. Guo, and T. Zhang, \href{https://doi.org/10.1016/j.physleta.2022.128133}{Phys. Lett. A \textbf{439}, 128133 (2022)}. 
	
	
	
	\bibitem{boyd} R. Boyd,  \textit{Nonlinear Optics}, 4th ed. (Academic, New York, 2020). 
	
	\bibitem{aliBellBBO} A. Motazedifard, S. A. Madani, and N. S. Vayaghan, 
	\href{https://doi.org/10.1007/s11082-021-03067-8}{Quantum Electron. \textbf{53}, 378 (2021)}.
	
	\bibitem{aliDNA} A. Motazedifard and S. A. Madani, 
	\href{https://doi.org/10.1364/OSAC.413830}{OSA Continuum \textbf{4}, 1049 (2021)}.
	
    \bibitem{FrequencyDependentSqueezing} J. Junker, D. Wilken, N. Johny, D. Steinmeyer, and M. Heurs, 
    \href{https://doi.org/10.1103/PhysRevLett.129.033602}{Phys. Rev. Lett. \textbf{129}, 033602 (2022)}.


	
	\bibitem{aliBellPPKTP} A. Motazedifard, S. A. Madani, J. J. Dashkasan, and N. S. Vayaghan, 
	\href{https://doi.org/10.1016/j.heliyon.2021.e07384}{Heliyon. 7 (6), e07384 (2021)}. 
	
	
	\bibitem{shySPDC} Y. Shih, 
	\href{https://doi.org/10.1088/0034-4885/66/6/203}{Rep. Prog. Phys. \textbf{66}, 1009 (2003)}.
	
	
	
	
	\bibitem{Takeno2007}Y. Takeno, M. Yukawa, H. Yonezawa, and A. Furusawa, \href{https://doi.org/10.1364/OE.15.004321}{Opt. Express \textbf{15}, 4321 (2007)}. 
	
	
	\bibitem{Dwyer2022}S.E. Dwyer, G. L. Mansell, and L. McCuller, \href{https://doi.org/10.3390/galaxies10020046}{Galaxies \textbf{10}, 46 (2022)}. 


	
	
\end{thebibliography}
\end{document}